\documentclass[final,5p,twocolumn]{elsarticle}

\usepackage{geometry}
\usepackage{graphicx}
\usepackage{amssymb}
\usepackage{amsmath}
\usepackage{multirow}
\usepackage{tabularx}
\usepackage{lipsum}
\usepackage{microtype}
\usepackage{color}
\usepackage[hyphens]{url}

\usepackage[T1]{fontenc}

\usepackage{fancyhdr}
\journal{Nuclear Instruments and Methods A}

\begin{document}

\begin{frontmatter}
 
  \title{CLARION2-TRINITY: a Compton-suppressed HPGe and GAGG:Ce-Si-Si array for absolute cross-section measurements with heavy ions}
 
  \author{T.~J.~Gray}
  \ead{graytj@ornl.gov}
  \author{J.~M.~Allmond}
  \author{D.~T.~Dowling}
  \author{M.~Febbraro}
  \author{T.~T.~King}
  \author{S.~D.~Pain}
  \author{D.~W.~Stracener}
  \address{Physics Division, Oak Ridge National Laboratory, Oak Ridge, Tennessee, USA}    
    
  \author{S.~Ajayi}
  \author{J.~Aragon}
  \author{L.~Baby}
  \author{P.~Barber}
  \author{C.~Benetti}
  \author{S.~Bhattacharya}
  \author{R.~Boisseau}
  \author{J.~Gibbons}
  \author{S.~L.~Tabor}
  \author{V.~Tripathi}
  \author{C.~Wibisono}
  \author{I.~Wiedenhoever}
  \address{Department of Physics, Florida State University, Tallahassee, Florida, 32306, USA}
    
  \author{L.~Bignell}
  \author{M.~S.~M.~Gerathy}
  \author{G.~Lane}
  \author{L.~J.~McKie}
  \author{A.~J.~Mitchell}
  \author{J.~Pope}
  \author{R.~du~Rietz}
  \author{A.~E.~Stuchbery}
  \address{Department of Nuclear Physics and Accelerator Applications, Research School of Physics, Australian National University, Canberra, 2601, ACT, Australia}
    
  \author{K.~Vaigneur}
  \address{Agile Technologies, Inc., 10337 Yellow Pine Ln, Knoxville, TN 37932, USA}  
  
  \author{T.~J.~Ruland}
  \address{Department of Physics and Astronomy, Louisiana State University, Baton Rouge, Louisiana 70803, USA}

  \begin{abstract}
    The design and performance of a new Compton-suppressed HPGe and charged-particle array, CLARION2-TRINITY, are described. The TRINITY charged-particle array is comprised of 64 Cerium-doped Gadolinium Aluminium Gallium Garnet (GAGG:Ce) crystals configured into five rings spanning 7--54 degrees, and two annular silicon detectors that can shadow or extend the angular coverage to backward angles with minimal $\gamma$-ray attenuation. GAGG:Ce is a non-hygroscopic, bright, and relatively fast scintillator with a light distribution well matched to SiPMs. Count rates up to 40~kHz per crystal are sustainable. Fundamental characteristics of GAGG:Ce are measured and presented, including light- and heavy-ion particle identification (PID) capability, pulse-height defects, radiation hardness, and emission spectra. The CLARION2 array consists of up to 16 Compton-suppressed HPGe Clover detectors ($\approx4\%$ efficiency at 1 MeV) configured into four rings (eight HPGe crystal rings) using a non-Archimedean geometry that suppresses back-to-back coincident 511-keV gamma rays. The entire array is instrumented with 100- and 500-MHz (14 bit) waveform digitizers which enable triggerless operation, pulse-shape discrimination, fast timing, and pileup correction. Finally, two examples of experimental data taken during the commissioning of the CLARION2-TRINITY system are given: a PID spectrum from $^{16}$O + $^{18}$O fusion-evaporation, and PID and Doppler-corrected $\gamma$-ray spectra from $^{48}$Ti + $^{12}$C Coulomb excitation.
  \end{abstract}
  
\end{frontmatter}

\section{Introduction}
CLARION2-TRINITY is a new particle-$\gamma$ coincidence detector array designed for absolute cross-section measurements with inverse-kinematic reactions, e.g., single-step Coulomb excitation and sub-barrier transfer. The system was largely inspired by the older CLARION-BAREBALL (and Microball) system \cite{Gross2000, GalindoUribarri2010, Sarantites1996} used at the Holified Radioactive Ion Beam Facility (HRIBF). CLARION2-TRINITY has many notable evolutionary steps including: new, flexible mechanical infrastructure that supports more detectors and beam heights; new digital electronics that enable triggerless operation and pulse-shape discrimination; and a new charged-particle detector that is faster and more granular. The system is envisioned to primarily operate at the John~D.~Fox Laboratory of Florida State University (FSU), the nuCARIBU-ATLAS facility of Argonne National Laboratory, and the ReA3 facility of FRIB-MSU.

\par%
The new charged-particle array TRINITY was designed to precisely measure heavy-ion induced absolute cross-sections at ``safe'' sub-barrier energies. Integrated beam currents or normalizations to absolute cross sections can be determined from Rutherford scattering relative to the inelastic or sub-barrier reactions, e.g., $\sigma_{\mathrm{Coulex}} = \sigma_{\mathrm{Rutherford}} \times P(I_i \rightarrow I_f)$, where elastic and inelastic contributions sum to the Rutherford cross-section. Furthermore, at sufficiently low beam energies, the heavy-ion induced reactions become completely independent of the optical potential, see Refs.~\cite{Allmond2011,Allmond2012,Allmond2014PRC,Allmond2014PRL,Stuchbery2013}. This ``Rutherford normalization'' can be achieved with high precision at backward center-of-mass scattering angles where elastic yields change slowly and are less sensitive to geometric uncertainties. For normal kinematics, when the beam is lighter than the target, a large fraction of the backward center-of-mass angles are contaminated with target shadows due to beam-like nucleus recoils near $90^{\circ}$ in the lab frame. This is an especially large problem for thick targets. However, in inverse kinematics, when the beam is heavier than the target, the recoiling beam and target nuclei generally travel forward in the lab frame. This means that the majority of the backward center-of-mass angles are void of target shadows and the lighter recoiling target nuclei, which are easier to measure and exhibit reduced pulse-height defects, carry most of the recoiling energy. In most practical scenarios with finite target thicknesses, 7--54 degrees in the lab frame captures the majority of target recoils. For lab angles greater than 54 degrees, one begins to lose the lower-energy recoiling target nuclei from reactions at the front of thick ($\approx 1-2$~mg/cm$^2$) targets.
\par%
In light of this intended useage, the following characteristics were considered in the charged-particle detector design (in no particular order):
\begin{enumerate}
\item \textit{Low thresholds for heavy ions} --- to maintain a clean Rutherford normalization in inverse kinematics for the lighter recoiling target nuclei at the largest lab angles.
\item \textit{Fast decaying signals with good time resolution} --- to reduce real and false pileups from weak or fluctuating ``slow'' signals. Pileup reduces the efficiency, limits the usable beam current, and requires corrections to the particle yields. Good time resolution also provides a reliable time reference for HPGe calibrations.
\item \textit{Intrinsic Particle IDentification (PID) capability} --- to provide selectivity to the Rutherford scattering and nucleon transfer channels.
\item \textit{Moderate energy resolution} --- to provide additional selectivity and diagnostics to expected or unexpected radiation with different energies.
\item \textit{Cylindrical or spherical symmetry} --- to match the symmetry of cross-section and angular-correlation calculations, Doppler corrections, and to mitigate uncertainties related to beam size and location.
\item \textit{Granularity} --- to provide good Doppler corrections, angular correlations, and sensitivity to center-of-mass scattering angles.
\item \textit{Durability, simplicity, and flexibility} --- to minimize maintenance and storage requirements, reduce setup time and complexity, and to accommodate supplemental detector additions for extended physics criteria.
\item \textit{Small total mass and size} --- to minimize $\gamma$-ray attenuation and HPGe radial distance.
\end{enumerate}
With these considerations, a relatively new scintillator --- Gadolinium Aluminium Gallium Garnet (Gd$_3$Al$_2$Ga$_3$O$_{12}$) doped with Ce (GAGG:Ce) --- was chosen and investigated as the basis for the new charged-particle array, TRINITY. GAGG:Ce is mechanically hard (8 out of 10 on the Mohs scale), radiation hard, high-Z ($\approx 54$), non-hygroscopic, fast ($\approx 92$ and \mbox{278-ns} decay components), 1.9 index of refraction at 540 nm, and bright ($\approx 46$,$000$ photons/MeV) with \mbox{$\approx 5\%$} energy resolution~\cite{Kamada2011, Kamada2011IEEE, Kamada2012, Kamada2012IEEE, Iwanowska2013, Tyagi2013, Kozlova2016, Stewart2016, Yoneyama2018}. Furthermore, it has intrinsic PID capability through two decay components with different decay times and varying relative amplitudes~\cite{Kim2015, Sibczynski2015, Rawat2016, Kobayashi2012, Tamagawa2015}. 
\par%
To investigate and further characterize GAGG:Ce as part of the development of the TRINITY array, several measurements with light and heavy ions were conducted at the Heavy Ion Accelerator Facility at the Australian National University (HIAF-ANU) and the John~D.~Fox Laboratory at Florida State University (FSU). New fundamental properties of GAGG:Ce are reported. The details of the TRINITY array itself will then be presented, along with details of the accompanying CLARION2 Clover array for $\gamma$-ray detection. Finally, performance results from TRINITY and CLARION2 commissioning experiments at FSU will be briefly shown.

\section{GAGG:Ce scintillator: prototyping and extended characterization}

\subsection{GAGG:Ce radioluminescence spectra}
\begin{figure}
  \centering
  \includegraphics[width=0.5\textwidth]{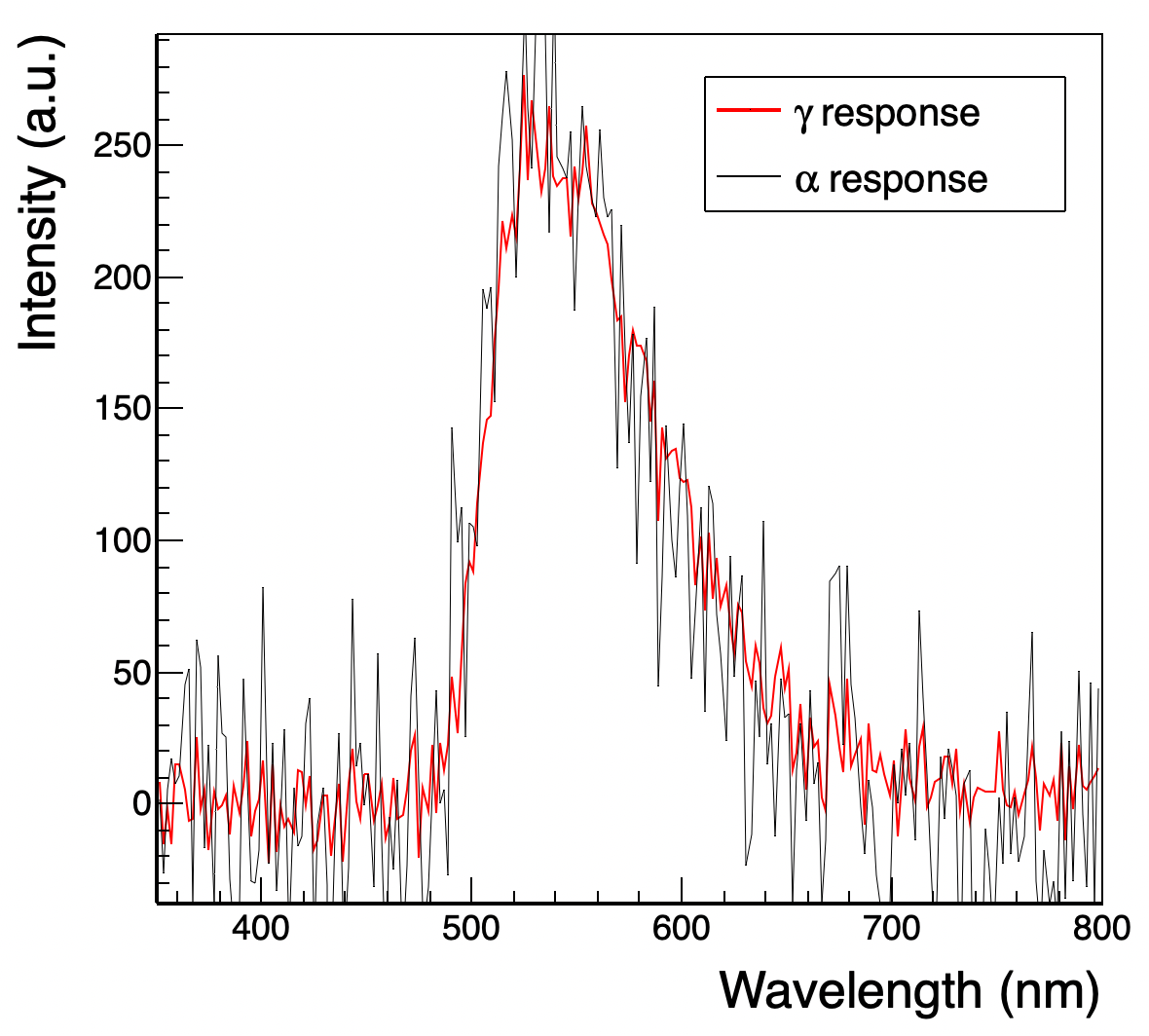}
  \caption{Radioluminescence spectra of a GAGG:Ce crystal from both $\alpha$ and $\gamma$ radiation.  Spectra were normalized by area. }
  \label{fig:gagg_emission}
\end{figure}

An important parameter for designing a detector based on scintillation is the emission spectrum, which dictates the choice of optical reflectors, waveguides, and photosensors for collection of the scintillation light. To address this, the wavelength-dependent radioluminescence spectrum for the prototype GAGG:Ce-LightGuide was measured for both $\gamma$-ray and $\alpha$-particle irradiations.  For the sources of $\gamma$ and $\alpha$ particles, $^{137}$Cs and $^{241}$Am sources were used, respectively. The radioluminescence spectra were recorded using a Horiba Fluoromax Plus spectrofluorometer. The results for both $\gamma$-ray and $\alpha$-particle irradiation are shown in Figure~\ref{fig:gagg_emission}, where the spectra are normalized by area. The two spectra are identical within the statistics of the measurement, with peak emission of 538~nm. This indicates there are no chromatic differences in the emission spectra for $\gamma$ rays and $\alpha$ particles, and thus a common spectrum can be used for the GAGG:Ce-LightGuide design.

\subsection{Detector preparation and readout}
\begin{figure}
  \centering
  \includegraphics[width=0.5\textwidth]{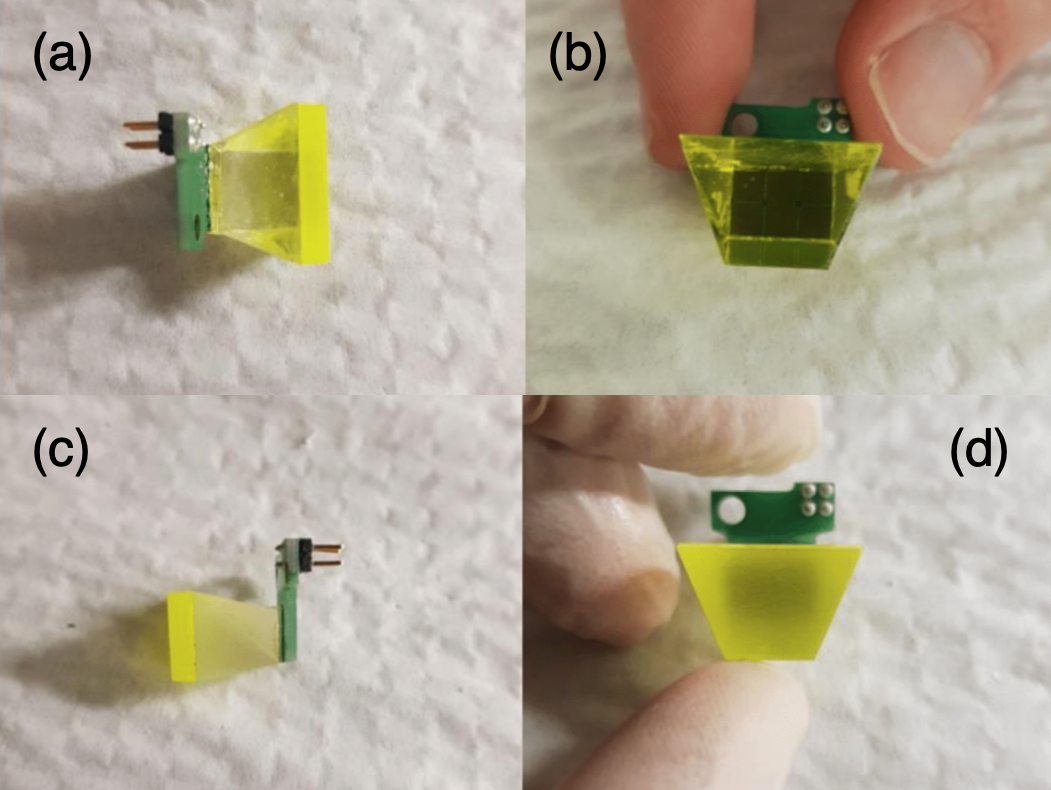}
  \caption{Photograph of the first prototype GAGG:Ce-LightGuide-SiPM detector: (a) and (b) show the prototype with specular polished surfaces, while (c) and (d) show the subsequent diffuse finish treatment, which gave $\approx 3$ times more light output than the former. }
  \label{fig:gagg_photo}
\end{figure}

Two prototype GAGG:Ce-LightGuide-SiPM detectors were fabricated and characterized to guide the final design. Figure~\ref{fig:gagg_photo} shows photographs of the first prototype. Initially, the surfaces were polished specular. However, after diffusing the surfaces, approximately three times more light yield was obtained. Thus, the final detectors have all side surfaces diffuse but the coupling surfaces were given a specular polish, including the front face to allow the possibility of adding a layer of slow plastic scintillator in the future. Yellow-emitting fast plastic films approximately 14-$\mu$m thick were developed at ORNL and coupled to the front face of the GAGG:Ce crystal. However, the combination of the relatively fast GAGG:Ce and the slow timing of the SiPM and associated circuitry resulted in no observable improvement to the PID. Each trapezoid-shaped GAGG:Ce crystal is $3$-mm thick and coupled to a rectangular segmented SiPM through a Borofloat glass light guide with EJ-500 optical cement~\cite{BOROFLOAT, EJ500}. The crystals and light guides were dry wrapped with 6.35-$\mu$m thick aluminized mylar reflector. Teflon tape was further dry wrapped around the light guides near the SiPM and held in place with kapton tape.

\par
\begin{figure}
  \centering
  \includegraphics[width=0.5\textwidth]{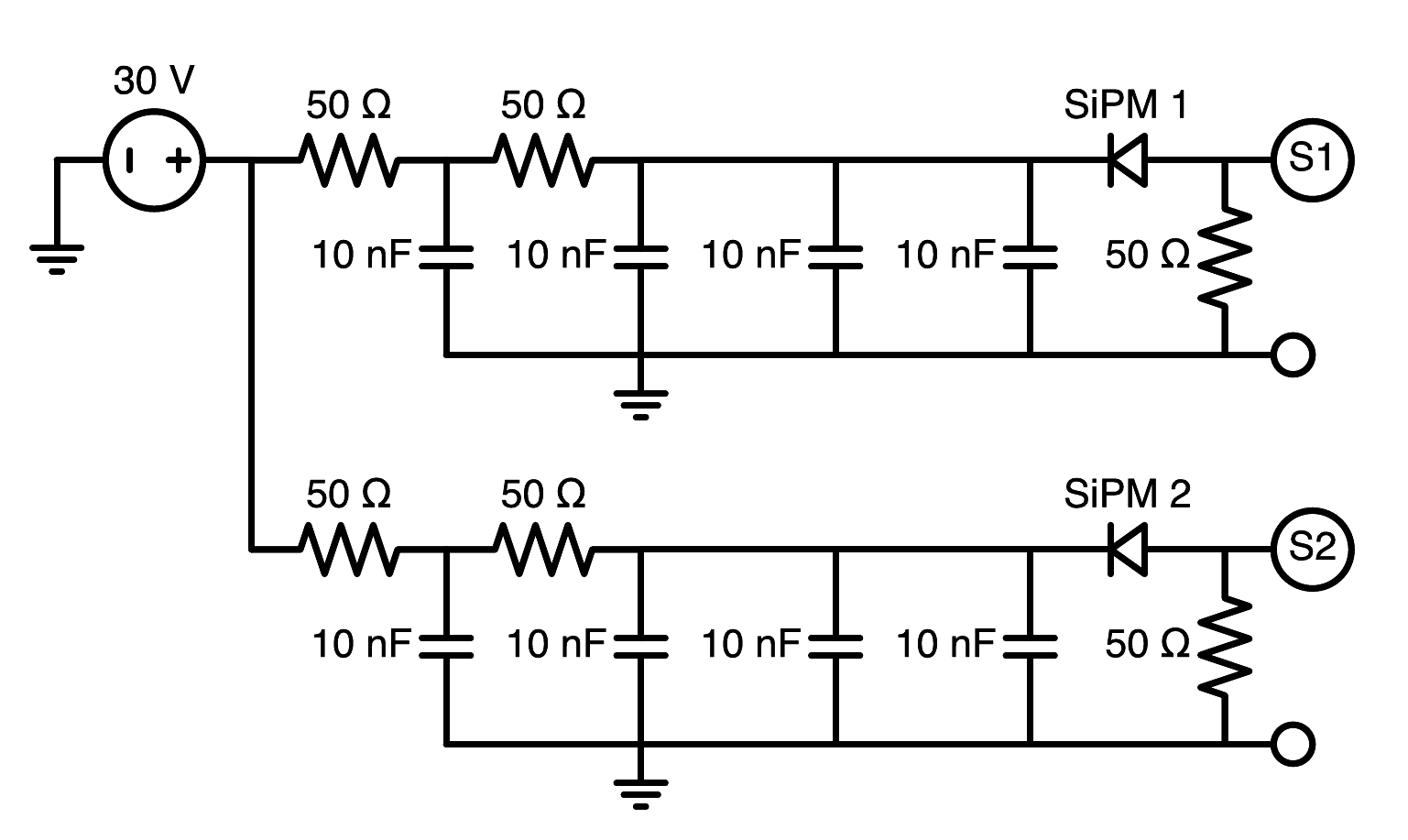}
  \caption{Conditioning circuit for the SiPM readout. Readout of the two SiPMs is taken from the contacts labeled ``S1'' and ``S2''.}
  \label{fig:condition}
\end{figure}

%
Ketek PM3315-WB Silicon photomultipliers (SiPM) are used as the photosensor. Each SiPM unit is $3 \times 3$~mm$^2$, has 38.8k cells with 15-$\mu$m pitch, $<1$-ns rise time, 13-ns recovery time, and 27-V breakdown voltage. The Photo Detection Efficiency (PDE) approaches roughly $22\%$ at 520~nm using an overvoltage of 5~V. ``Type 1'' SiPM assemblies have 233k total pixels, spread across 6 sensors in a $2\times3$ arrangement. ``Type 2'' SiPM assemblies have 310k total pixels in a $2 \times 4$ arrangement. The dynamic range is strongly influenced by the pixel count and recovery time. In both cases, the outputs of each row are bussed together to give two output signals per crystal. By collecting both signals and requiring a coincidence between the two, noise is greatly suppressed and thresholds can be set much lower. Each SiPM board is connected to a readout board using the conditioning circuit shown in Fig.~\ref{fig:condition} to provide filtered $+$30-V bias and a 50-$\Omega$ voltage divider that converts the current to an impedance-matched voltage signal. The conditioning circuit leads to an effective recovery time of 60~ns, which could be reduced by a future update to the readout boards. 

\subsection{GAGG:Ce characterization with light and heavy ions}
The literature on GAGG:Ce is primarily limited to characterization with $\gamma$-ray and $\alpha$ sources. There is a general lack of light-quenching and pulse-height defect (PHD) information with light and heavy ions, which is needed for developing and calibrating a charged-particle detector. Therefore, GAGG:Ce was studied systematically with $\gamma$-rays and charged particles ranging from protons to titanium.
\begin{figure}
  \centering
  \includegraphics[width=0.5\textwidth]{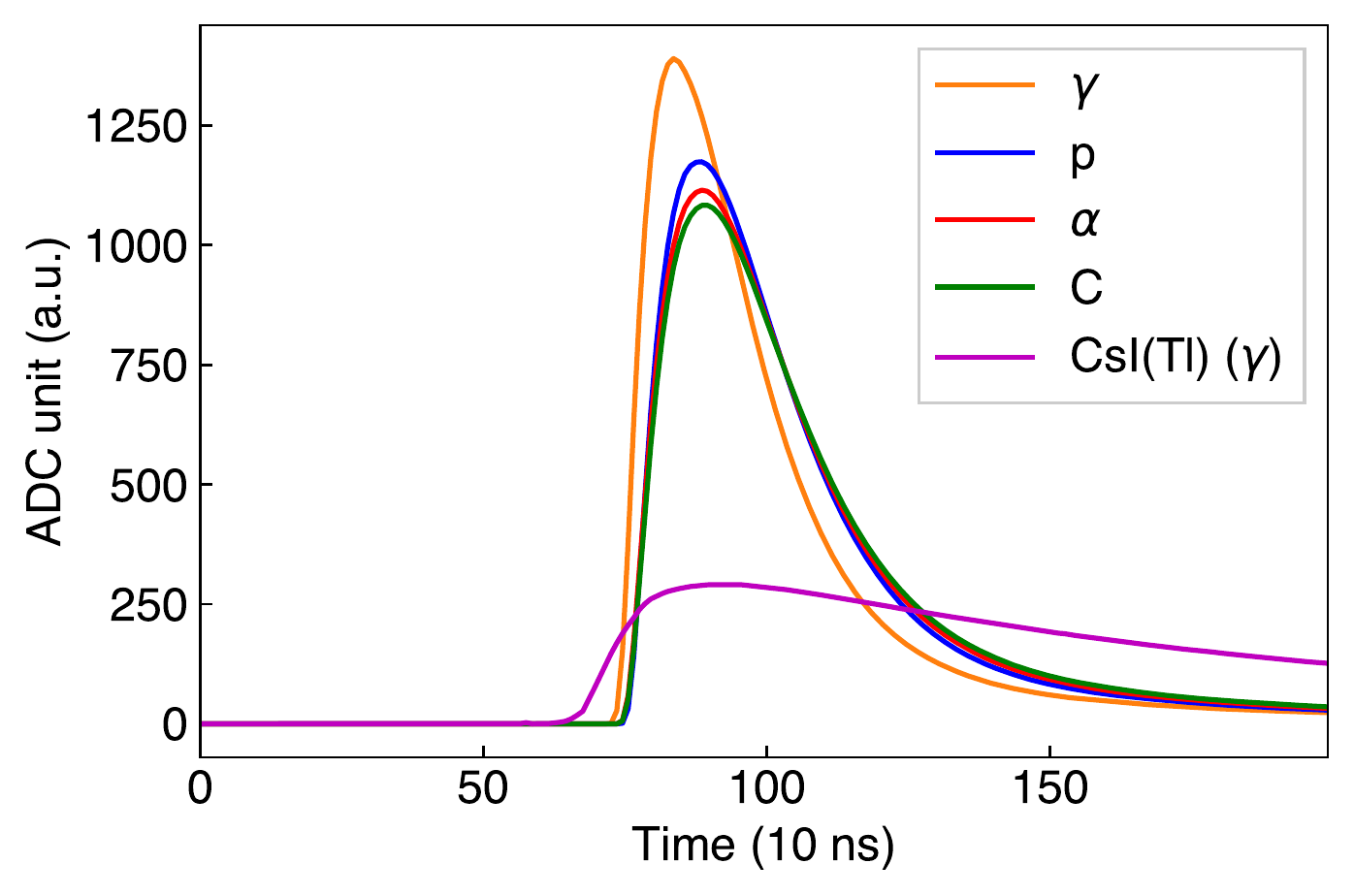}
  \caption{Super-pulses from proton, $\alpha$, and carbon charged particles in a GAGG:Ce crystal, formed from the sum of many individual waveforms. Each has been normalised to the same area; the different peak (fast) to tail (slow) ratios give PID capability. Note that $\approx 60$~ns response from the SiPM readout circuit is convoluted with the intrinsic GAGG:Ce pulse to give the above waveforms. The response of a CsI(Tl) crystal is given for comparison, see text.}
  \label{fig:superpulse}
\end{figure}
\par%
Pulse shapes for 662-keV $\gamma$ rays, 8-MeV protons, 15-MeV $\alpha$ particles, and 48-MeV carbon are shown in Fig.~\ref{fig:superpulse}. Each ``superpulse'' waveform has been constructed from the sum of many events and the individual ``superpulses'' have been re-normalised to the same area. The PID capability arises from the difference in shapes: $\gamma$-rays have more light output concentrated directly after the interaction, while protons, $\alpha$ particles, and carbon have progressively larger delayed components as compared to the fast component. The shape of a typical CsI(Tl) response to $\gamma$ rays with the same SiPM and readout is included for reference --- note the much longer rise time, decay time, and reduced signal height; CsI(Tl) and GAGG:Ce are comparable in brightness but with different decay times.
\par
\begin{figure}
  \centering
  \includegraphics[width=0.5\textwidth]{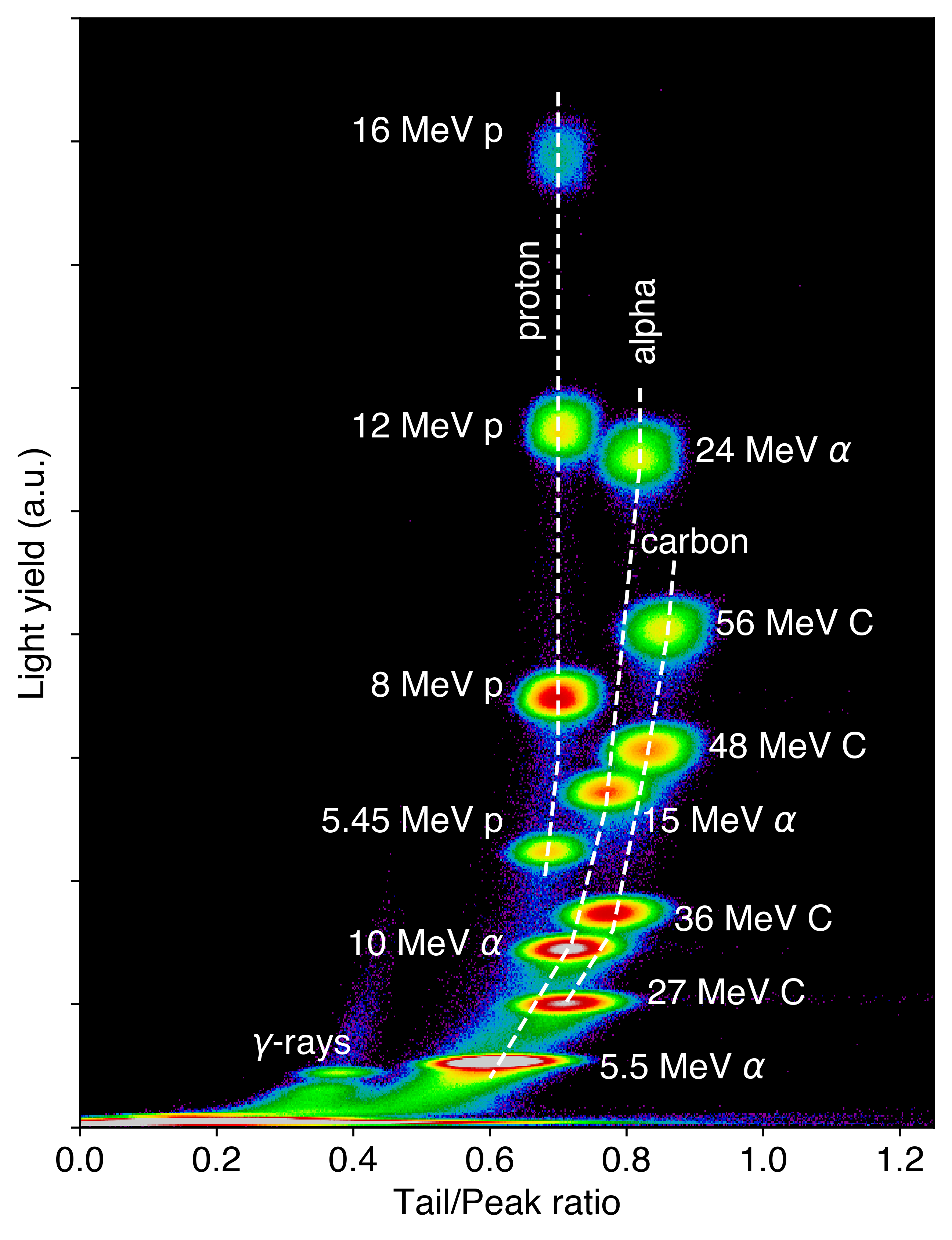}
  \caption{PID from forward scattering on a thin Au foil. White dashed lines to guide the eye connect the proton, $\alpha$, and carbon peaks.}
  \label{fig:pid_evsr}
\end{figure}

%
The PIDs of forward scattering proton, $\alpha$, and $^{12}$C beams on a $0.2$~mg/cm$^{2}$ Au foil are shown in Fig.~\ref{fig:pid_evsr}. The PID was obtained by comparing waveform integrals (baseline corrected) of the fast ``peak'' and delayed ``tail'' components of the GAGG:Ce signal. The ratio of these two components becomes more pronounced and linearized with increasing energy. There is a clear difference in the PID for 8-MeV protons, 15-MeV $\alpha$ particles, and 48-MeV $^{12}$C nuclei, despite all these giving a total light yield (i.e., trace or waveform integral) which varies by only $\approx 20\%$.
\par
\begin{figure}
  \centering
  \includegraphics[width=0.5\textwidth]{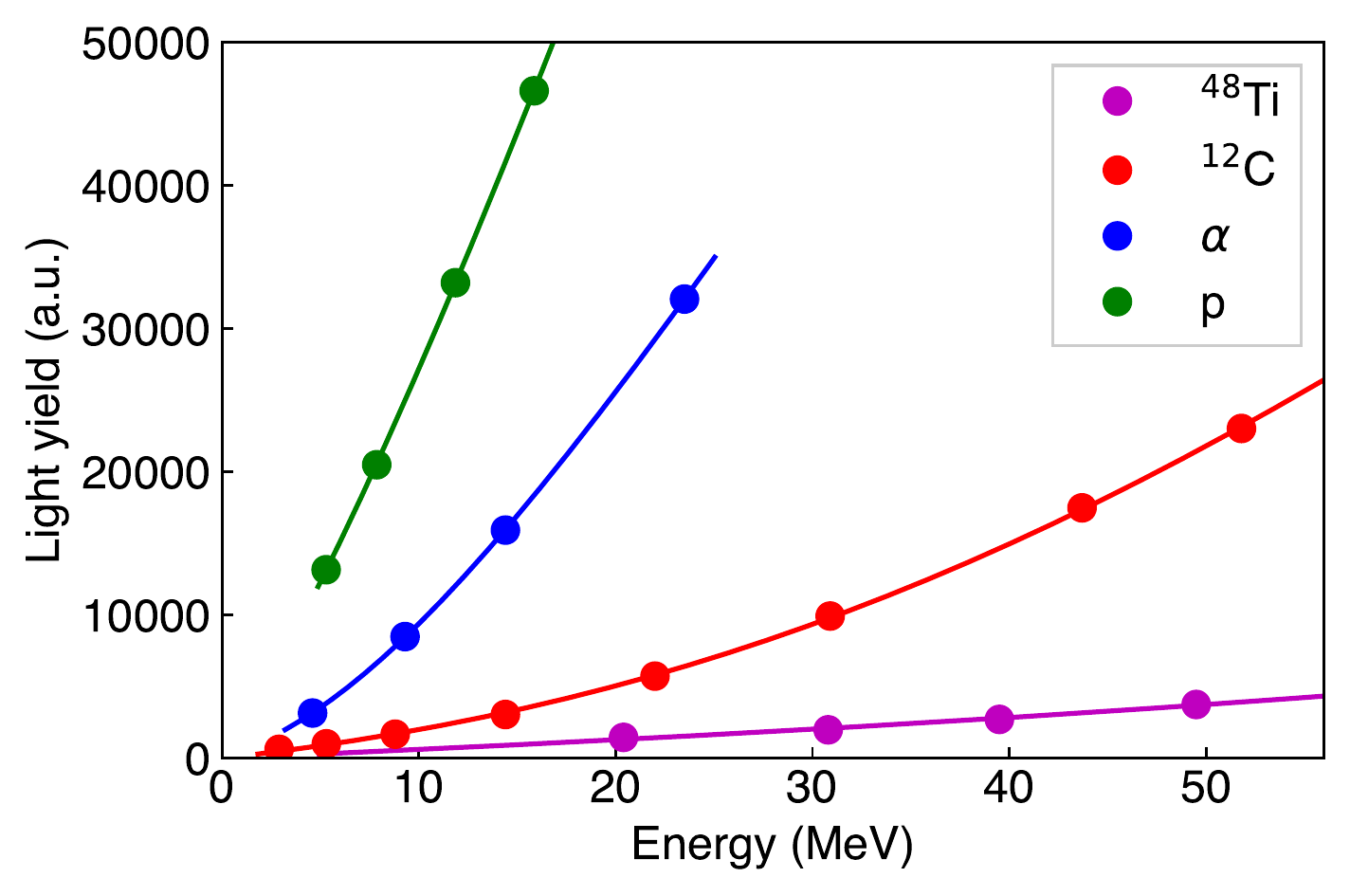}
  \caption{Nonlinearity and pulse-height defect of the GAGG:Ce response at low energies. The energies were corrected for kinematics and energy loss through the aluminized mylar foil to provide the energy seen by the GAGG:Ce crystal.}
  \label{fig:nonlin}
\end{figure}

\begin{figure}
  \centering
  \includegraphics[width=0.5\textwidth]{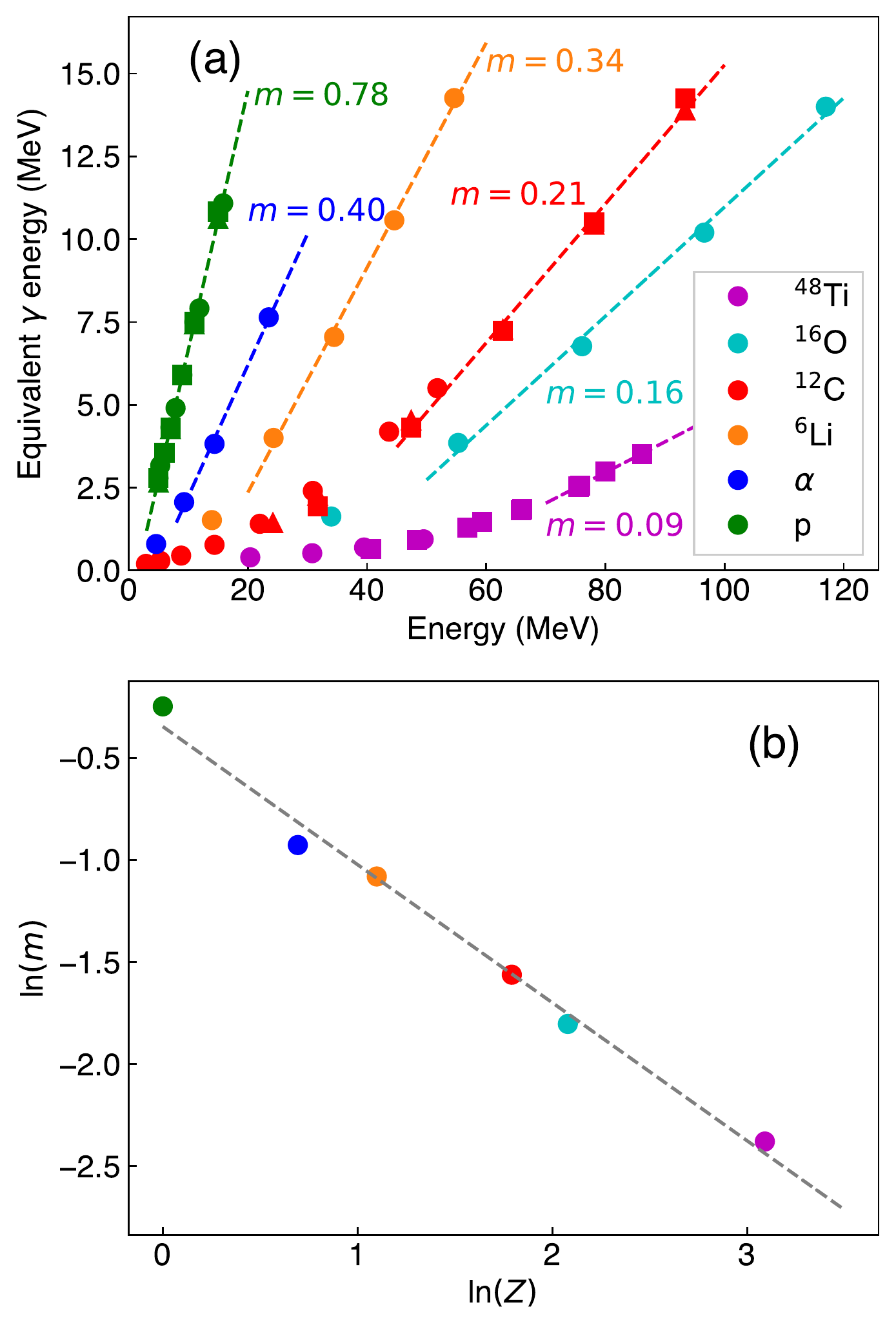}
  \caption{(a) Light output in equivalent $\gamma$-ray energy plotted for various charged particles. Note the light output is non-linear for  $< 2$~MeV equivalent $\gamma$ energy. Several data-sets taken under different conditions (different crystal prototypes) are combined for proton, carbon, and Ti curves; these agree well. (b) shows a linearized fit of $\ln(Z)$ vs $\ln(m)$ where $m$ are the gradients fitted in (a). This provides a phenomenological model for the pulse-height defect,
$\approx 0.71/Z^{0.68}$.}
  \label{fig:phd}
\end{figure}

%
The nonlinearity and pulse-height defect of the GAGG:Ce response at low energies is shown in Fig.~\ref{fig:nonlin} for proton, $\alpha$, $^{12}$C, and $^{48}$Ti beams. The energies were corrected for kinematics and energy loss through the aluminized mylar foil to provide the energy seen by the GAGG:Ce crystal. A 10-MeV $^{12}$C particle induces a much weaker signal than a 10-MeV $\alpha$ or proton. This is quantified in Fig.~\ref{fig:phd}, where the energy of the incident charged particle is plotted against the equivalent $\gamma$-ray energy in terms of GAGG:Ce light output, i.e., trace integral. A fit to the linear region of each species gives the pulse height defects; the gradient for the fit corresponds to $\text{Light yield}(\gamma)/\text{Light yield} (\text{charged particle})$. The relationship between the pulse height defects and $Z$ is shown in Table~\ref{tab:phd} and Fig.~\ref{fig:phd} (b). A power law ($\mathrm{PHD} \approx 0.71/Z^{0.68}$) describes the data well. It seems that the GAGG:Ce response is linear when the light output is equivalent to a $\gamma$-ray energy of $\approx 2$~MeV or higher. 

\begin{table}
  \centering
  \caption{Pulse-height defects for proton, $\alpha$, $^{7}$Li, 
    $^{12}$C, and $^{16}$O in GAGG:Ce, defined as the gradient of the linear region in Fig.~\ref{fig:phd} (a). }
  \label{tab:phd}
\begin{tabularx}{\columnwidth} { 
    >{\centering\arraybackslash}p{1.8cm}
    >{\centering\arraybackslash}p{1.8cm}
    >{\centering\arraybackslash}p{1.8cm}
    >{\centering\arraybackslash}X
  }
                      & PHD  & $1/Z$ & $0.71/Z^{0.68}$ \\ \hline
      $\gamma/\gamma$ & 1.0  &       &                 \\
      $\gamma/$p      & 0.78 & 1     & 0.71                  \\
      $\gamma/\alpha$ & 0.40 & 0.50  & 0.44                 \\
      $\gamma/$Li     & 0.34 & 0.33  & 0.34                 \\
      $\gamma/$C      & 0.21 & 0.16  & 0.21                 \\
      $\gamma/$O      & 0.16 & 0.125 & 0.17  \\
      $\gamma/$Ti     & 0.09 & 0.045 & 0.09 \\
   \end{tabularx}
\end{table}

\par
\begin{figure}
  \centering
  \includegraphics[width=0.5\textwidth]{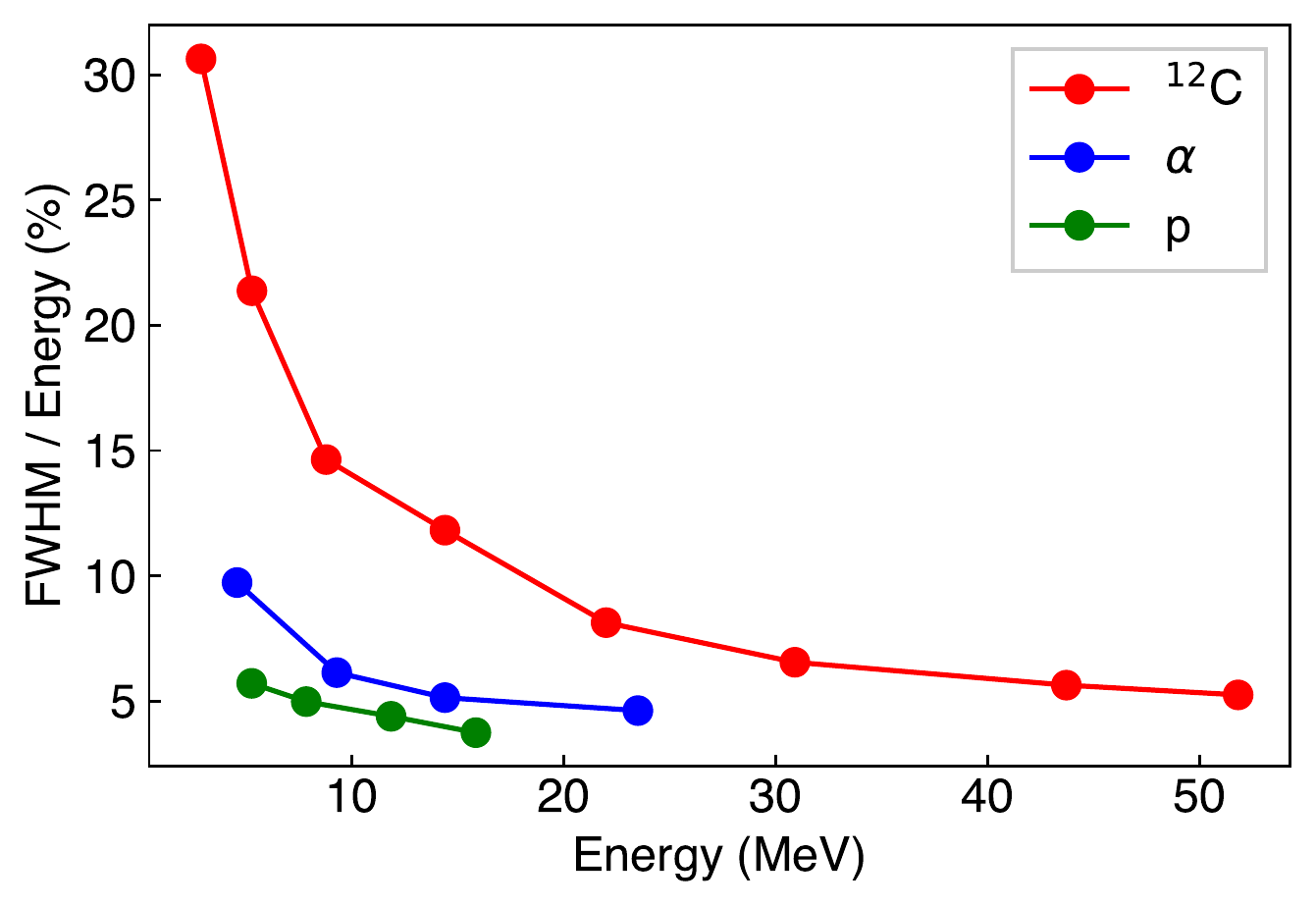}
  \caption{The resolution of GAGG:Ce crystals for carbon, alpha, and proton ions as a function of energy. The heavier ions approach $\approx 5\%$ at high energies while the protons reached $3.7\%$.}
  \label{fig:resolution}
\end{figure}

%
The resolution of the GAGG:Ce crystals for several energies and radiation types is shown in Fig.~\ref{fig:resolution}. The resolution approaches $\text{FWHM}/\text{Energy} \approx 5\%$ as energy increases for the heavier ions while the protons reached $3.7\%$.

\par
\begin{figure}
  \centering
  \includegraphics[width=0.5\textwidth]{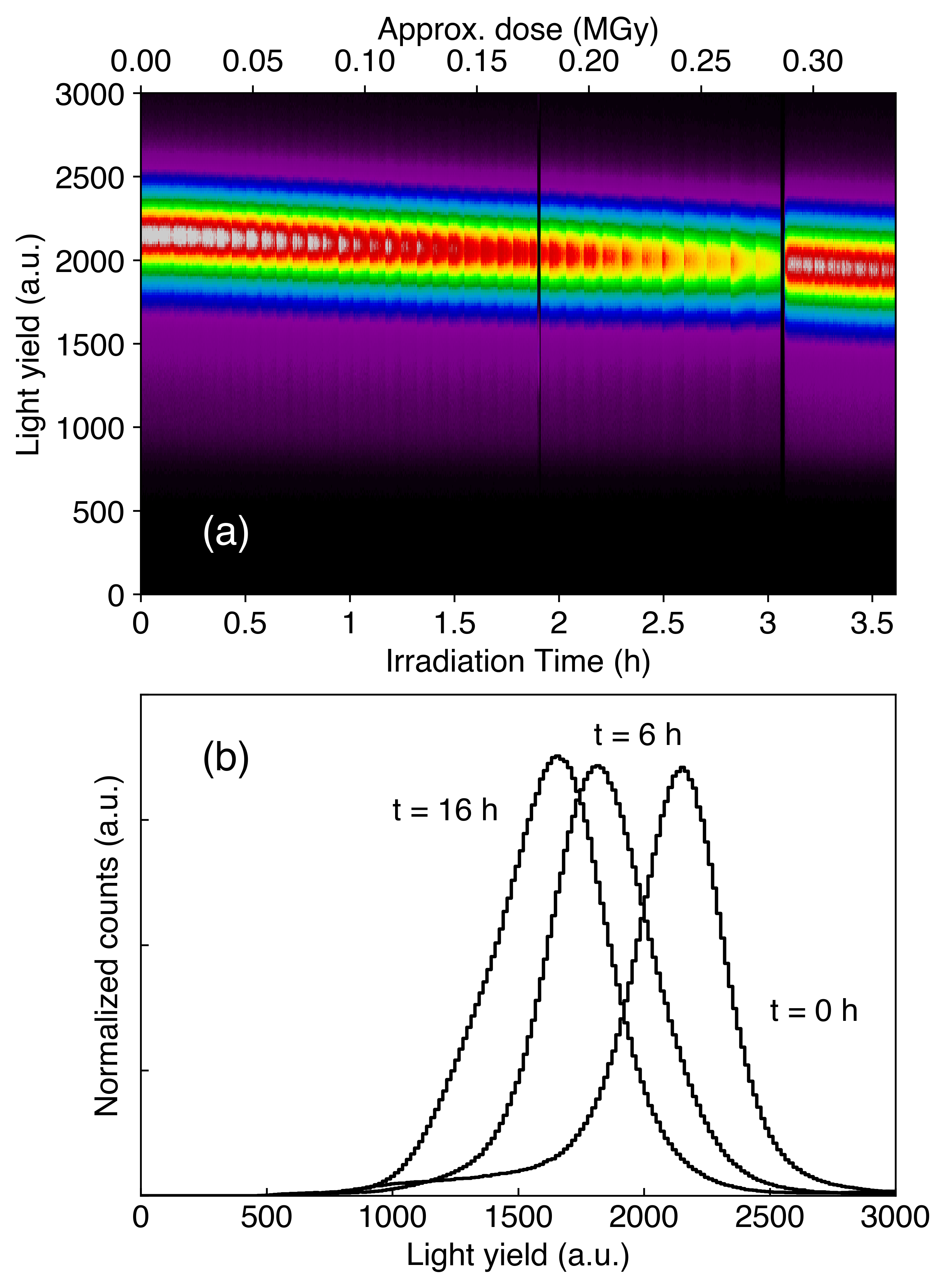}
  \caption{Radiation damage on a small spot on one GAGG:Ce crystal from a
    75.8-MeV $^{50}$Ti beam after passing through a C target. Light
    yield drops to $\approx 75\%$ of the original value after 16 hours. The flux is estimated as $1.5 \times 10^{7}$ particles per second per
    cm$^{2}$. For 16 hours this corresponds to $\approx 3.5 \times
    10^{9}$ particles. The black vertical lines in panel (a) correspond to beam interruptions, while the non-uniform pattern corresponds to beam intensity fluctuations.}
  \label{fig:rad_dmg}
\end{figure}

%
The effects of heavy-ion induced radiation damage were also investigated. A 75.8-MeV $^{50}$Ti beam was incident on a $0.6$~mg/cm$^2$ $^{12}$C target with a collimated GAGG:Ce detector placed downstream at zero degrees. The energy loss through the C target was measured as 19.3~MeV, meaning that 56.5~MeV was the average $^{50}$Ti energy incident on the GAGG:Ce crystal. The collimator was a thick sheet of tantalum metal (sufficient to stop the beam) with a small $0.7$-mm diameter hole. The GAGG:Ce counted at $\approx 60$~kHz on this small concentrated spot over the course of 16 hours with minimal interruptions in beam delivery. This corresponds to a flux of $\approx1.5 \times 10^{7}$ particles per second per cm$^{2}$. A drop to $\approx 75\%$ light output was observed with a total of $\approx 3.5 \times 10^{9}$ particles, $\approx 3.2 \times 10^{-2}$~J, or $\approx 1.4$~MGy integrated dose, assuming that the $^{50}$Ti stops in the first 8.9~$\mu$m of GAGG:Ce. The first $\approx 4$ hours are shown in Fig.~\ref{fig:rad_dmg}~(a). Figure~\ref{fig:rad_dmg}~(b) shows several snapshots of this data: immediately after data collection started, after 6 hours, and after 16 hours. The energy resolution was also degraded by the accumulated radiation damage. Note that the flux of the direct beam on the zero-degree detector used to produce Fig.~\ref{fig:rad_dmg} is several orders of magnitude larger than the scattered beam on a regular uncollimated crystal during the course of an experiment. To reach an equivalent dose corresponding to a $25\%$ drop in light output would require continuous operation at a count rate of 30~kHz per crystal for two years.

\section{TRINITY array}
\begin{figure}
  \centering
  \includegraphics[angle=0,width=0.4\textwidth]{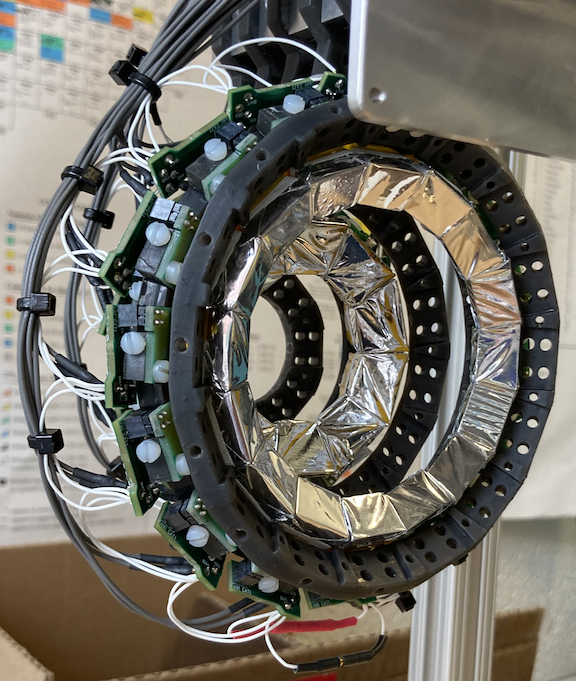}
  \caption{Photograph of the TRINITY array with rings 2 and 4.}
  \label{fig:trinity}
\end{figure}

\begin{figure}
  \centering
  \includegraphics[angle=0,width=0.4\textwidth]{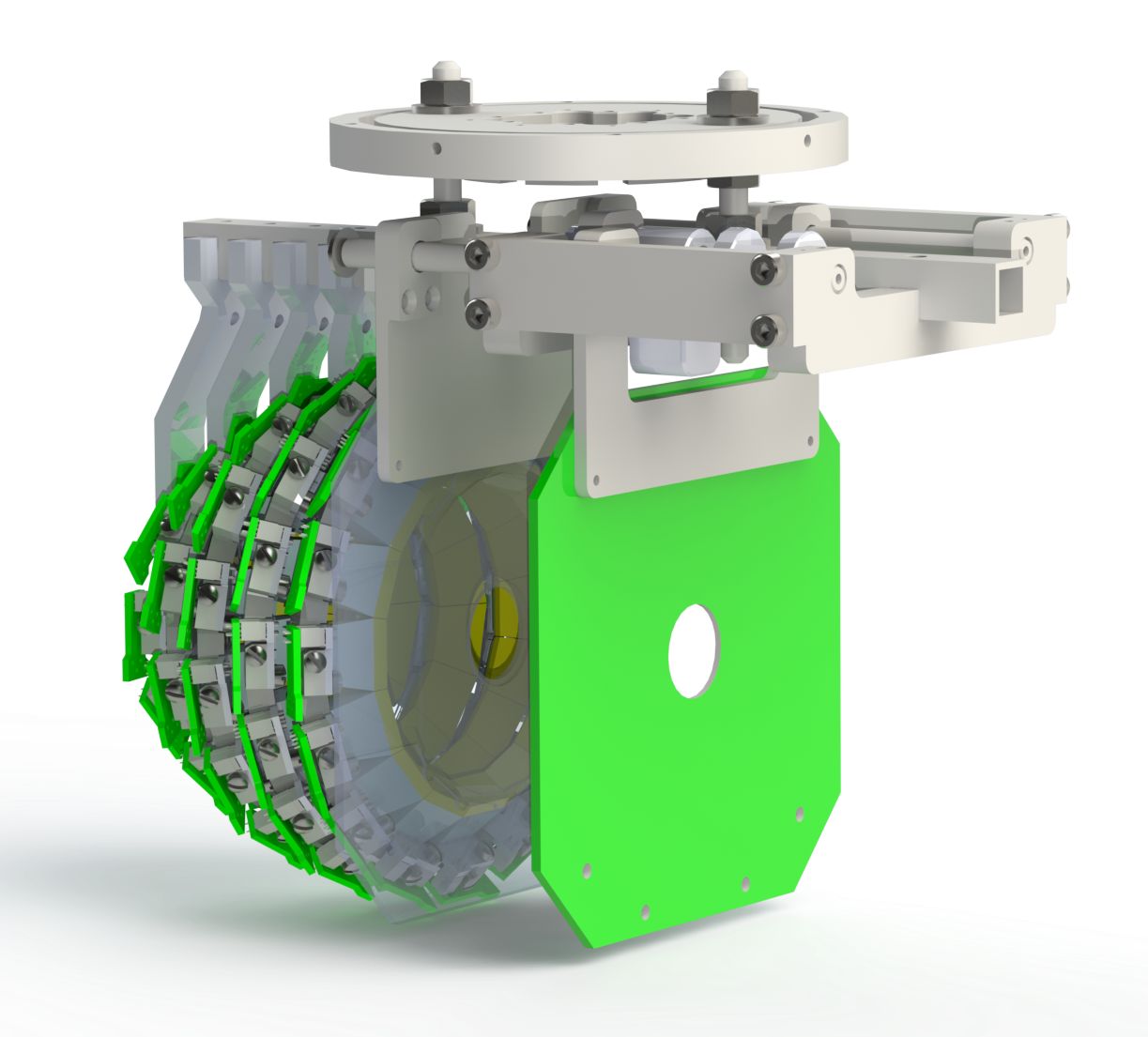}
  \caption{Rendering of the full TRINITY array with GAGG:Ce crystals and two annular Si detectors. One Si detector is rendered in green, while the second is semi-transparent and just in front of the GAGG:Ce crystals.}
  \label{fig:trinity_render}
\end{figure}

The TRINITY charged-particle array is comprised of three main parts --- two annular S2/S3 Si detectors from Micron Semiconductor Ltd (forward and/or backward lab angles), and a ball of GAGG:Ce crystals at forward lab angles of 7--54 degrees, which corresponds to the majority of recoiling target nuclei in most inverse kinematics experiments, but only $20.2\%$ of full $4\pi$ coverage. The crystals are arranged in five rings subtending $\approx 10^{\circ}$ each, with full $\phi$ coverage. The details of the $\theta$ angles and number of crystals for each ring are in Table~\ref{tab:trinity}. Each ring has the last crystal centered vertically (i.e. $\phi = 0$). The crystals are 3~mm thick, and have a trapezoidal geometry with areas that range between $\approx 2-3.5$~cm$^{2}$ from ring to ring. A photograph of the TRINITY frame with rings 2 and 4 populated is in Fig.~\ref{fig:trinity}, and a computer rendering of TRINITY with all five rings and Si detectors is shown in Fig.~\ref{fig:trinity_render}.

\begin{table}
  \centering
  \caption{Angles and number of crystals for each GAGG:Ce ring. Each ring
  has the last crystal centered on $\phi = 0^{\circ}$.}
  \label{tab:trinity}
\begin{tabularx}{\columnwidth} { 
  >{\centering\arraybackslash}p{1.0cm}
  >{\centering\arraybackslash}p{1.5cm}
  >{\centering\arraybackslash}p{1.5cm}
  >{\centering\arraybackslash}p{1.5cm}
  >{\centering\arraybackslash}X
}
Ring & $\theta_{\mathrm{low}}$ ($^{\circ}$) & $\theta_{\mathrm{high}}$ ($^{\circ}$) & crystals & SiPM type \\ \hline
1    & 7.0                           & 14.0                     & 8     &  1 \\
2    & 14.0                          & 24.0                     & 10    &  2 \\
3    & 24.0                          & 34.0                     & 14    &  2 \\
4    & 34.0                          & 44.0                     & 16    &  2 \\
5    & 44.0                          & 54.0                     & 16    &  2 \\
                                           
\end{tabularx}
\end{table}

\par%
The S2/S3 detectors can be mounted on one or both of the sliding plate fixtures, enabling one to independently adjust the forward and backward laboratory angle coverage of the silicon detectors. For the forward lab angles,  the silicon can be placed to shadow the GAGG:Ce rings to act as a $\Delta$E detector, or extend the angular coverage to wider scattering angles than the GAGG:Ce coverage. The S2 detectors have 16 sectors and 48 rings with 0.491-mm pitch. The S3 detectors have 32 sectors and 24 rings with 0.886-mm pitch. Both S2 and S3 detectors have similar form factors with active inner/outer radii of 22~mm and 70~mm, respectively, and PCB inner/outer radii of 20~mm and 76~mm, respectively. Thicknesses are typically 1~mm or less.
\par%
The forward silicon detector can be continuously positioned from the target plane to 24.5 mm downstream with Ring 5 installed; removal of Ring 5 allows positions extending up to 38.0 mm. Positioning the silicon at 8~mm from the target plane results in 54$^{\circ}$--77$^{\circ}$ coverage. At the 24.5-mm position, the silicon subtends 24$^{\circ}$--55$^{\circ}$ (the angular coverage of GAGG:Ce Rings 3--5). With Ring 5 removed, placing the silicon at 38~mm gives 16$^{\circ}$--43$^{\circ}$ coverage (almost complete coverage of Rings 1--4). The upstream silicon detector can be positioned continuously between 16.8~mm and 78.7~mm from the target plane, corresponding to 116$^{\circ}$--147$^{\circ}$ coverage and 156$^{\circ}$--172$^{\circ}$, respectively. The chamber is sufficiently large to support charged foils or permanent magnets to suppress $\delta$ rays and, therefore, mitigate leakage current.
\par%
The silicon detectors are instrumented with charge-sensitive preamplifiers of the LASSA design, implemented in 36-channel preamplifier motherboards as used for ORRUBA \cite{Pain_ORRUBA_NIM} and GODDESS \cite{Pain_PhysProc_GODDESS_2017, Pain_Stardust_2014}. Signals are brought out of vacuum via custom multi-pin electrical feedthroughs. The preamplifier motherboards capacitively couple the detector to the preamplifiers, provide power and distribute detector bias voltage. The preamplifiers have a decay time constant of $\approx 50$~$\mu$s, and are available in two gains, nominally 15~mV/MeV and 60~mV/MeV. 

\subsection{Zero-degree detector}
A detector at zero degrees is useful for several applications including measurements of beam composition and empirical measurements of the energy loss through the target (or equivalently, stopping powers when combined with the target thickness); integrated cross-sections to first order are only sensitive to the total energy loss. Currently a single GAGG:Ce crystal is installed downstream of the target at zero degrees for this purpose, which is sufficient for stable beams with isotope contamination with different energies. A continuous plunger is located directly upstream of the zero-degree GAGG:Ce crystal, which can block or collimate the beam onto different spots of the zero-degree detector. During nominal operation the zero-degree detector is blocked completely to prevent radiation damage from the direct beam. In the future, a fast ionization chamber will be built and installed for use with radioactive beams containing isobar contamination at equivalent energies. This detector will be of the form developed for ORRUBA \cite{Chae_NIM_2014_FastIC} and upgraded for GODDESS \cite{Cheetham_thesis}, consisting of a series of 10-cm grids of 0.018-mm diameter gold-plated tungsten wires, oriented in planes normal to the beam direction, with 12.7-mm grid spacing in a stack alternating between anode and cathode grids. The 12.7-mm maximum electron drift distance enables fast counting, and the grids can be combined in groups for measurement of $\Delta E$ and residual energy. Such detectors can achieve particle identification of $\Delta Z=1$ up to about $A=50$, and counting rates exceeding 500k ions/second. Particle identification up to about $A=130$ can be achieved with lower rates. These detectors are read out by the same charge-sensitive preamplifiers as used for the silicon, with the decay time modified to be shorter (12 or 25 $\mu$s) to manage electronic pileup at high incident rates.

\par
\begin{figure}
  \centering
  \includegraphics[width=0.5\textwidth]{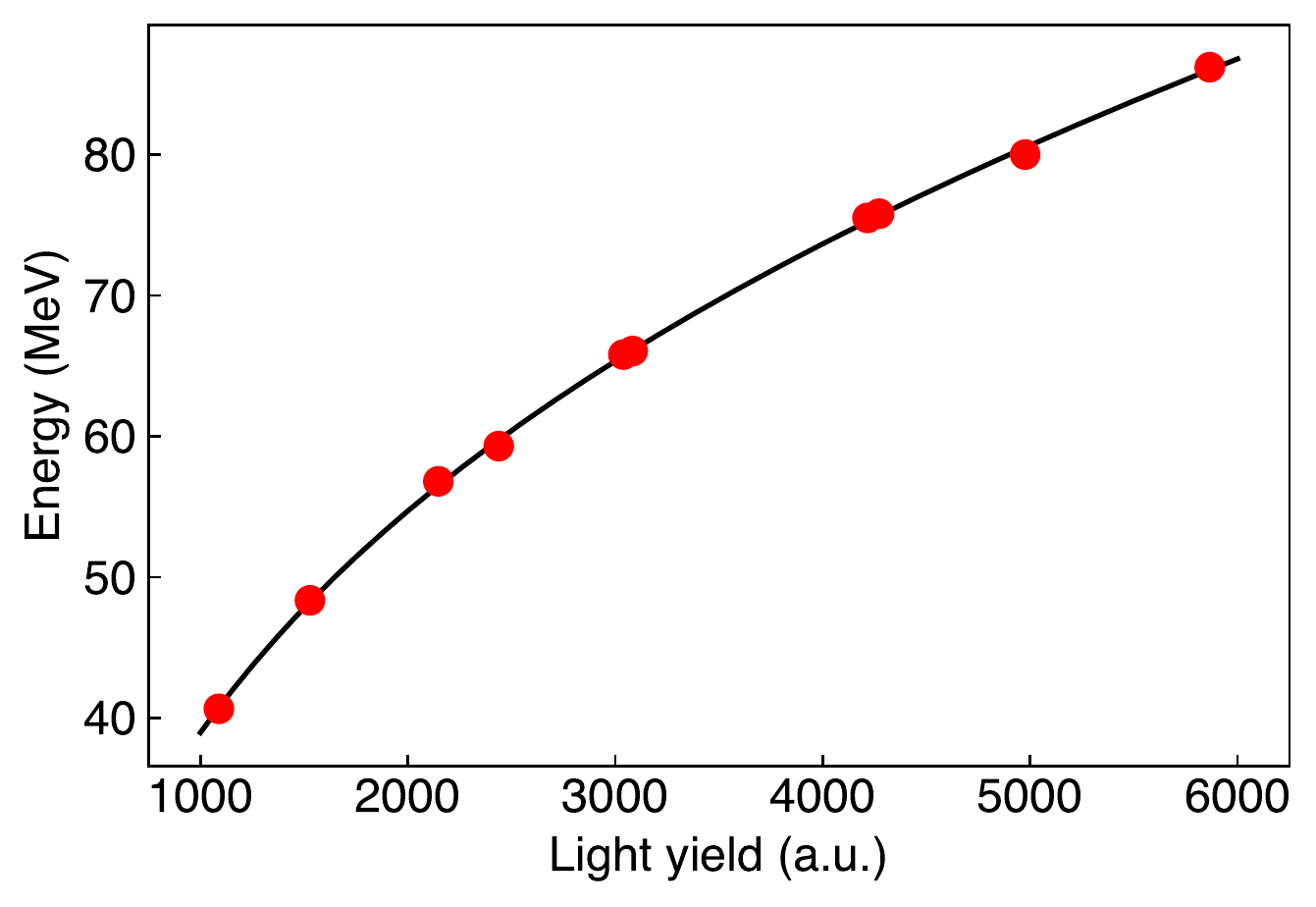}
  \caption{Ti calibration of the zero-degree GAGG:Ce detector for beam composition and energy loss measurements.}
  \label{fig:zero_cal}
\end{figure}

\begin{figure}
  \centering
  \includegraphics[width=0.5\textwidth]{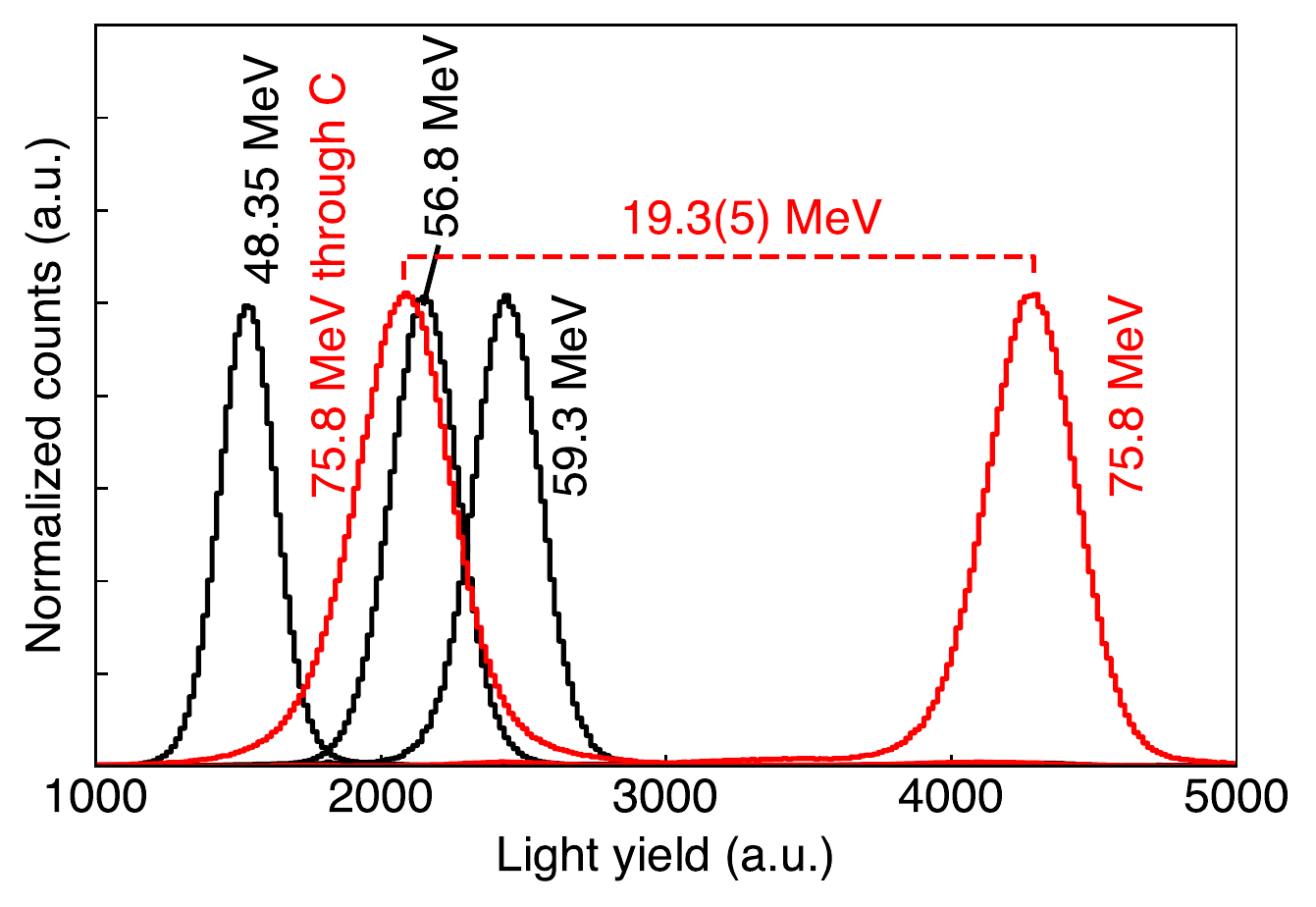}
  \caption{Example energy loss measurement with Ti-beam calibration points (black) and two 75.8-MeV beam measurements (red): one measurement with the C target in place, and one without.}
  \label{fig:stopping}
\end{figure}

%
An example of calibrating the zero-degree GAGG:Ce detector for energy loss measurements is shown in Fig.~\ref{fig:zero_cal}. Several Ti beam energies are incident directly on the zero-degree detector with no target in place. A minimum of three calibration points is necessary to account for nonlinearities. After this, a measurement is taken with the target in place. This allows the energy loss of the Ti beam through the C target to be measured empirically --- this is an important quantity for absolute cross-section measurements by Rutherford normalization, especially for thick targets. Some of the spectra used are shown in Fig.~\ref{fig:stopping}. The black spectra correspond to calibration points at different energies without the target in place, and the red corresponds to a 75.8-MeV beam, with and without the 0.6-mg/cm$^{2}$ $^{12}$C target in place.

\par
\begin{figure}
  \centering
  \includegraphics[width=0.5\textwidth]{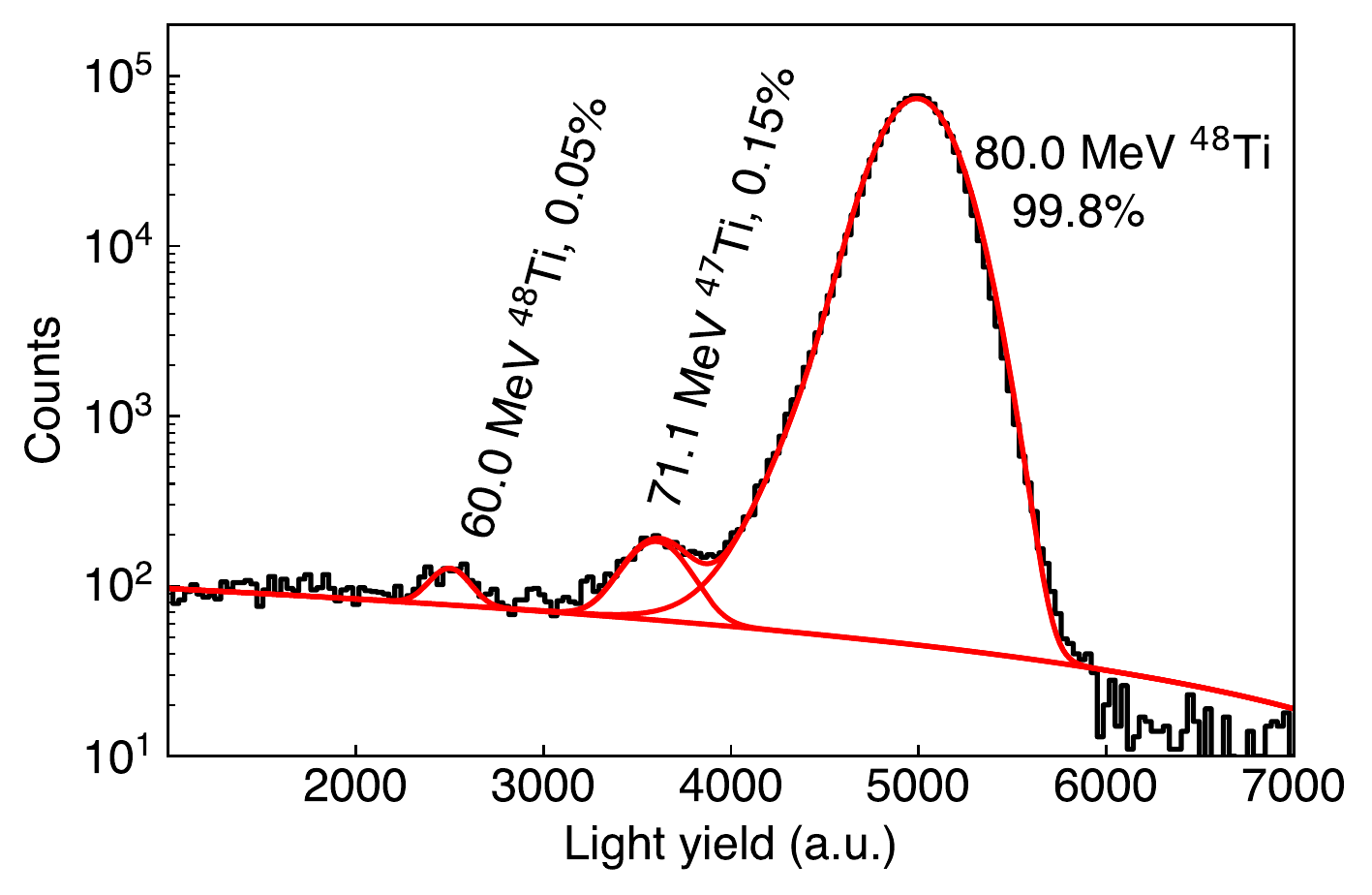}
  \caption{Beam composition measurement. The two minor components of
    the beam at 60.0 MeV and 71.1 MeV can be quantified.}
  \label{fig:beam_comp}
\end{figure}

%
The zero-degree GAGG:Ce detector can also be used to measure beam contamination, provided it is at a different energy to the dominant beam species. Figure~\ref{fig:beam_comp} shows this: while 99.8$\%$ of the beam is 80.0~MeV $^{48}$Ti, two minor components: 60.0-MeV $^{48}$Ti and 71.1-MeV $^{47}$Ti can be quantified. The contaminants are present in the beam due to double stripping and different charge-state combinations giving very similar magnetic rigidity. They are identified by this constraint along with their energy.

\section{CLARION2 array}
\begin{figure*}
  \centering
  \includegraphics[angle=0,width=0.9\textwidth]{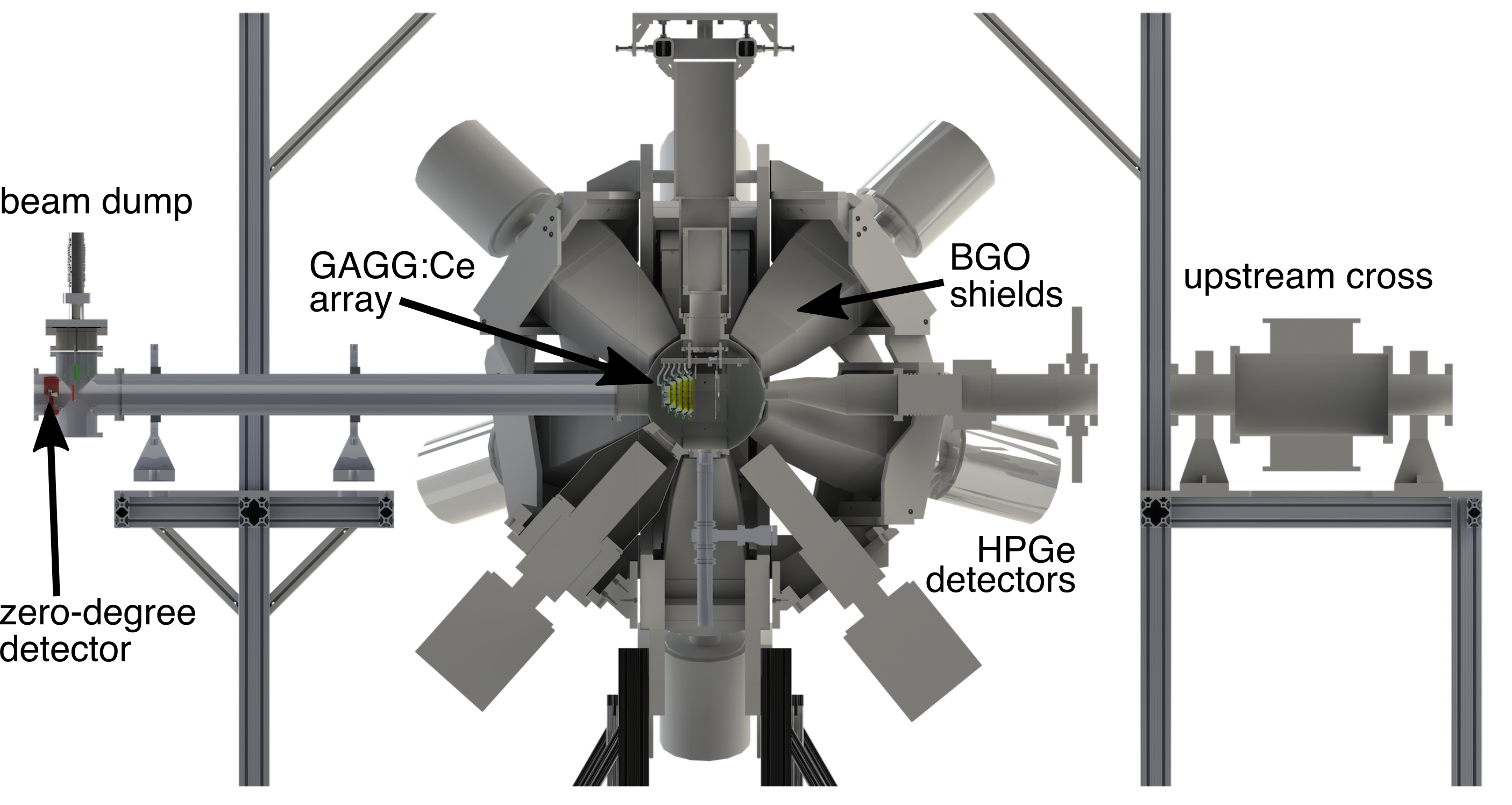}
  \caption{Computer rendering of a cross-section of the CLARION2-TRINITY frame, array, and beam line.}
  \label{fig:clarion_render}
\end{figure*}

\begin{figure}
  \centering
  \includegraphics[width=0.5\textwidth]{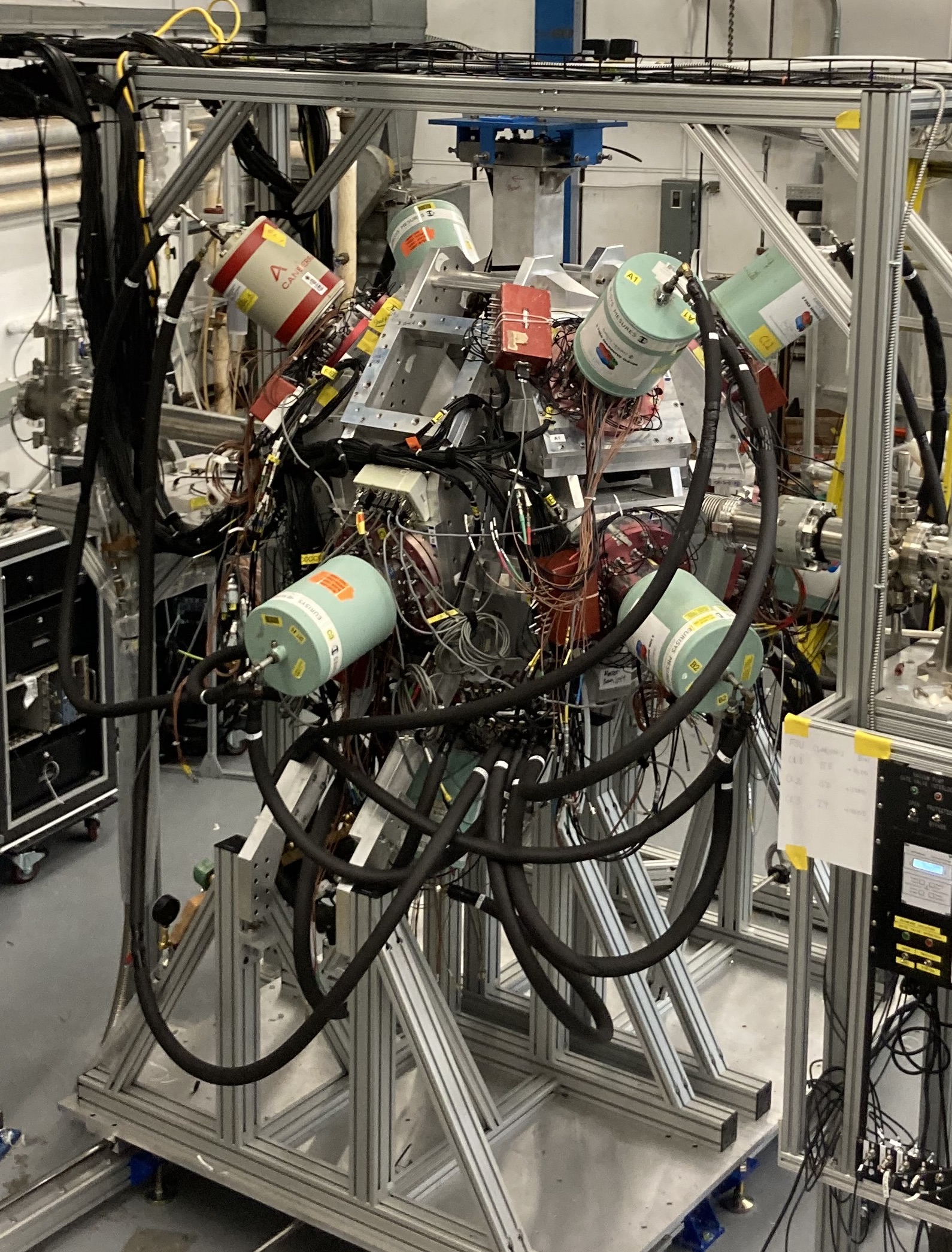}
  \caption{The CLARION2 array as configured during the first commissioning experiment at FSU.}
  \label{fig:clarion2}
\end{figure}

The CLARION2 $\gamma$-ray array has 16 ports suitable for Scionix BGO shields and EURISYS MESURES Clover HPGe detectors. Currently 11 of the ports (A1 -- K11) are populated with BGO shields and 9 of those have Clover detectors (1, 2, 3, 4, 5, 7, 8, 9, and 11) installed at the time of commissioning. The five remaining ports ($\alpha 1 - \alpha 3, \beta 1 - \beta 2$) can support additional Compton-suppressed Clover detectors or 4-way clusters of LaBr$_3$ for fast-timing measurements. A computer rendering of a cross-section of the CLARION2 and TRINITY arrays, beam pipe, frame, and downstream zero-degree detector is shown in Fig.~\ref{fig:clarion_render}. A photograph of the array as configured during the first commissioning experiment at FSU is show in Fig.~\ref{fig:clarion2}.

\par
\begin{figure}
  \centering
  \includegraphics[width=0.5\textwidth]{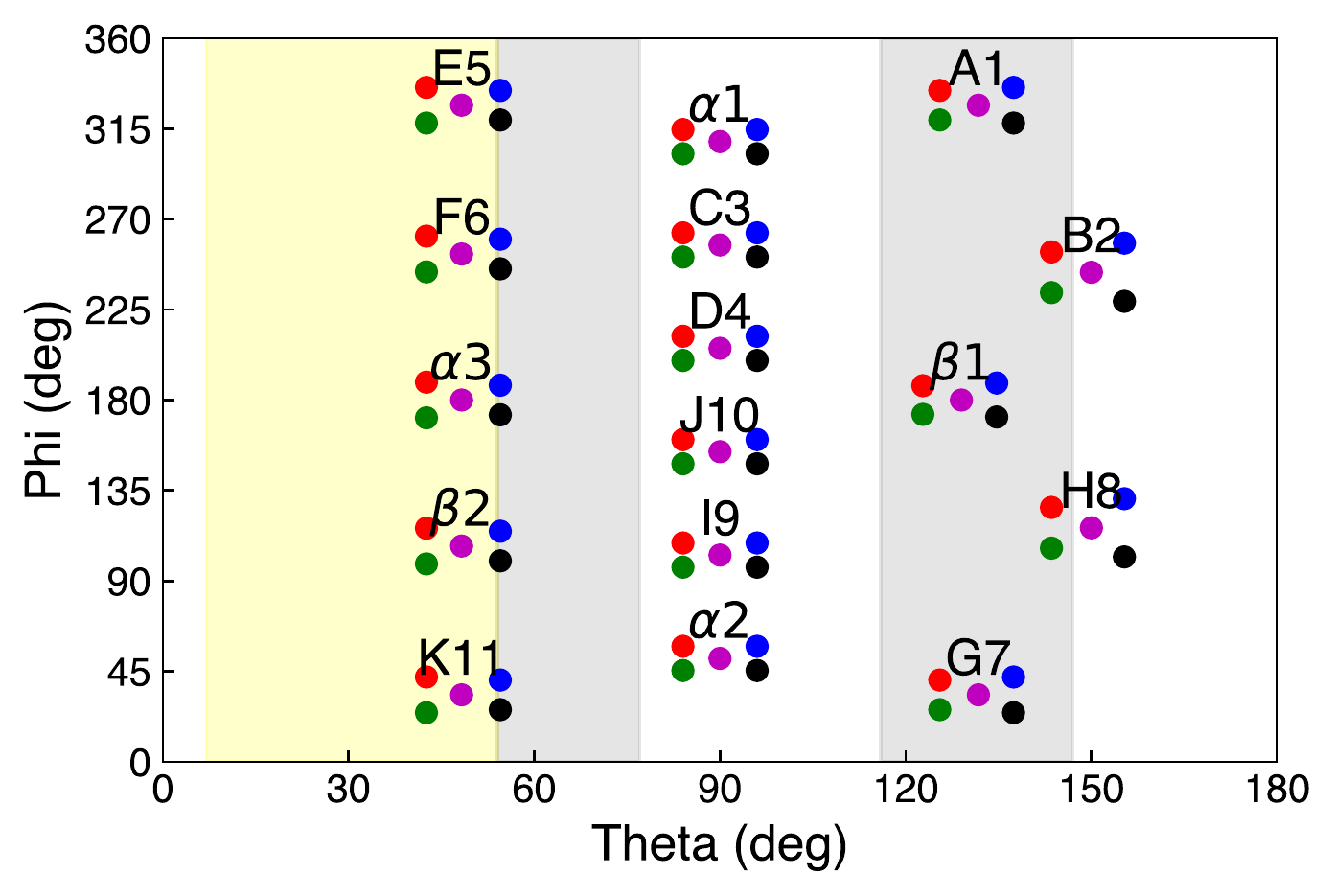}
  \caption{Angle map for CLARION2. The crystal angles represent the
    most inward position, where crystal faces are $20$~cm from the target position. The yellow shaded region from 7--54 degrees represents the GAGG:Ce angular coverage, while the two grey shaded regions show the angular coverage of the annular Si detectors in their nominal position; see text for further details.}
  \label{fig:clar_map}
\end{figure}

%
The radial distance of each Clover detector can be set to one of three positions inside the BGO shields: the innermost with the crystal faces $r = 20.0$~cm from the target position, and the others with $r = 21.75, 23.5$~cm, respectively. The BGO shields have 20-mm thick Hevimet collimators on the front face, which is at $r \approx 15$~cm from the target position.
While the CLARION2 frame was designed to accommodate a variety of BGO shields and Clovers from within the US community, the nominal design was formed around the Clover detectors and shields used in the CLARION array~\cite{Gross2000}. These Clovers have n-type crystals with starting geometry of 5-cm diameter and 8-cm length, before flattening two sides for close arrangement. The four individual crystals are segmented longitudinally into two halves, leading to four individual crystal signals, and three (left/middle/right) position signals. The left signal corresponds to blue and black crystals, while the right signal corresponds to green and red crystals. Clovers are oriented such that the blue and black crystals are on the high $\theta$ side. This configuration means that the position information obtained from the left/middle/right signals corresponds to improved $\theta$ resolution. The angles of each port are given in Table~\ref{tab:clar_ang}, both the Clover center angle (which is independent of $r$), and the crystal face center positions when the Clover is in the innermost position with $r=20.0$~cm. These data are plotted in Fig.~\ref{fig:clar_map}.
\par
\begin{figure}
  \centering
  \includegraphics[width=0.5\textwidth]{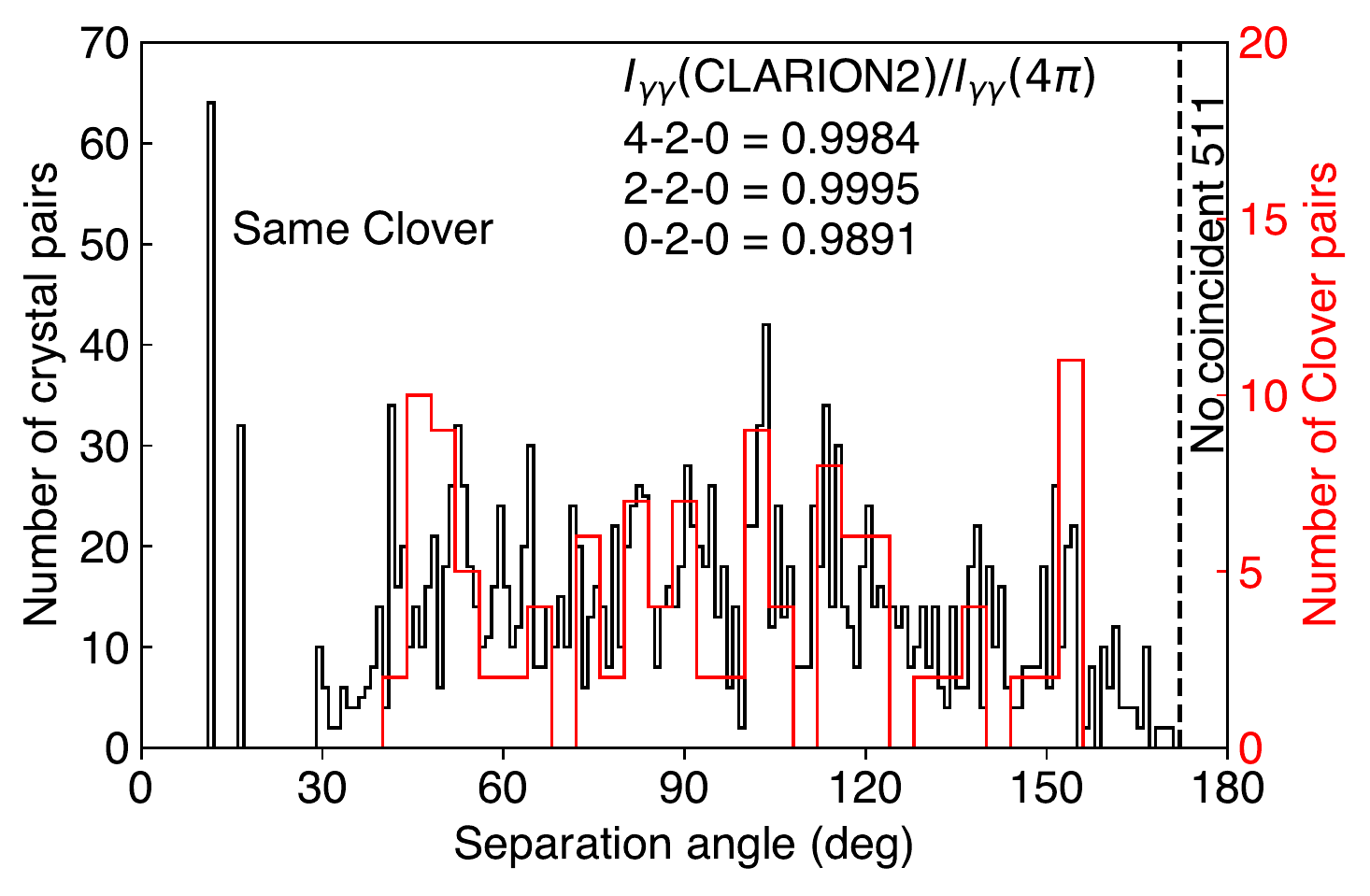}
  \caption{Histogram of the separation angles of every pair of
    crystals in CLARION2 (black), and separation angles between pairs of Clover centers (red). Even coverage from 30$^{\circ}$ to
    160$^{\circ}$ makes the system ideal for measuring
    $\gamma$-$\gamma$ angular correlations, while the lack of
    detectors with $\Delta \theta > 170$ means no coincident 511-keV
    $\gamma$-rays are observed. Correction coefficients for 4-2-0, 2-2-0, and 0-2-0 $\gamma$-$\gamma$ cascades are given for a fully populated array.}
  \label{fig:clar_ggang}
\end{figure}

%
There are six ports with $\theta = 90^{\circ}$, five with $\theta = 48.25^{\circ}$, two with $\theta = 131.75^{\circ}$, two with $\theta = 150^{\circ}$, and one with $\theta = 129^{\circ}$. It is worth noting that the geometry is non-Archimedian, and arranged such that no detectors have a separation of $180^{\circ}$. This means that no coincident 511-keV $\gamma$ rays are observed from pair production. A histogram of all $\gamma$-$\gamma$ angle combinations is shown in Fig.~\ref{fig:clar_ggang}. The black curve corresponds to angles between crystal centers, while the red corresponds to angles between Clover centers. The angular coverage is relatively even from $30^{\circ} - 170^{\circ}$, and there are no detector pairs with $\Delta \theta = 180^{\circ}$. The coincidence intensities of the array relative to a $4\pi$ detector are perturbed due to angular correlations. For common 4-2-0, 2-2-0, and 0-2-0 $E2$ $\gamma$-$\gamma$ cascades, this leads to ratios of $I_{\gamma \gamma}(\text{CLARION2})/I_{\gamma \gamma}(4 \pi) = 0.9984, 0.9995$, and $0.9891$, respectively for the fully populated array. The Clover positions were chosen to optimize these ratios and to eliminate coincident $511$-keV gamma rays.
\begin{table*}[t]
%  \centering
  \caption{CLARION2 slot names and positions. Crystal angles are valid for the closest Clover position, where crystal faces are $20$~cm from the target position.}
  \label{tab:clar_ang}
\begin{tabularx}{\textwidth} { 
    >{\centering\arraybackslash}p{0.7cm}
    >{\centering\arraybackslash}p{0.7cm}
    >{\centering\arraybackslash}p{1.2cm}
    >{\centering\arraybackslash}p{1.2cm}
    >{\centering\arraybackslash}p{1.2cm}
    >{\centering\arraybackslash}p{1.2cm}
    >{\centering\arraybackslash}p{1.2cm}
    >{\centering\arraybackslash}p{1.2cm}
    >{\centering\arraybackslash}p{1.2cm}
    >{\centering\arraybackslash}p{1.2cm}
    >{\centering\arraybackslash}p{1.2cm}
    >{\centering\arraybackslash}p{1.2cm}
  >{\centering\arraybackslash}X
}
Slot & Name                  & \multicolumn{2}{c}{Clover Center} & \multicolumn{2}{c}{Blue} & \multicolumn{2}{c}{Black} & \multicolumn{2}{c}{Green} & \multicolumn{2}{c}{Red}                                                                                         \\
   &  & $\theta$ ($^{\circ}$) & $\phi$ ($^{\circ}$)               & $\theta$ ($^{\circ}$)    & $\phi$ ($^{\circ}$)       & $\theta$ ($^{\circ}$)     & $\phi$ ($^{\circ}$) & $\theta$ ($^{\circ}$) & $\phi$ ($^{\circ}$) & $\theta$ ($^{\circ}$) & $\phi$ ($^{\circ}$) \\ \hline
1    & A1                    & 131.75                            & 326.69       & 137.43   & 335.56  & 137.43   & 317.82  & 125.50   & 319.32  & 125.50   & 334.05    \\
2    & B2                    & 150.00                            & 243.57       & 155.33   & 258.05  & 155.33   & 229.10  & 143.55   & 233.46  & 143.55   & 253.69    \\
3    & C3                    & 90.00                             & 257.14       & 95.99   & 263.17  & 95.99   & 251.12  & 84.01   & 251.12  & 84.01   & 263.17    \\    
4    & D4                    & 90.00                             & 205.71       & 95.99   & 211.74  & 95.99   & 199.69  & 84.01   & 199.69  & 84.01   & 211.74    \\    
5    & E5                    & 48.25                             & 326.68       & 54.50   & 334.05  & 54.50   & 319.32  & 42.57   & 317.82  & 42.57   & 335.56    \\    
6    & F6                    & 48.25                             & 252.64       & 54.50   & 260.01  & 54.50   & 245.28  & 42.57   & 243.77  & 42.57   & 261.52    \\    
7    & G7                    & 131.75                            & 33.31        & 137.43   & 42.18  & 137.43   & 24.44  & 125.50   & 25.95  & 125.50   & 40.68    \\     
8    & H8                    & 150.00                            & 116.42       & 155.33   & 130.90  & 155.33   & 101.95  & 143.55   & 106.31  & 143.55   & 126.54    \\
9    & I9                    & 90.00                             & 102.85       & 95.99   & 108.88  & 95.99   & 96.83  & 84.01   & 96.83  & 84.01   & 108.88    \\      
10   & J10                   & 90.00                             & 154.28       & 95.99   & 160.31  & 95.99   & 148.26  & 84.01   & 148.26  & 84.01   & 160.31    \\    
11   & K11                   & 48.25                             & 33.31        & 54.50   & 40.68  & 54.50   & 25.95  & 42.57   & 24.44  & 42.57   & 42.18    \\        
12   & $\alpha$1             & 90.00                             & 308.57       & 95.99   & 314.59  & 95.99   & 302.55  & 84.01   & 302.55  & 84.01   & 314.59    \\    
13   & $\alpha$2             & 90.00                             & 51.42        & 95.99   & 57.45  & 95.99   & 45.41  & 84.01   & 45.41  & 84.01   & 57.45    \\        
14   & $\alpha$3             & 48.25                             & 180.00       & 54.50   & 187.36  & 54.50   & 172.64  & 42.57   & 171.13  & 42.57   & 188.87    \\    
15   & $\beta$1              & 129.00                            & 180.00       & 134.71   & 188.44  & 134.71   & 171.56  & 122.77   & 172.87  & 122.77   & 187.13    \\
16   & $\beta$2              & 48.25                             & 107.35       & 54.50   & 114.72  & 54.50   & 99.99  & 42.57   & 98.48  & 42.57   & 116.23    \\  
\end{tabularx}
\end{table*}

\begin{figure}
  \centering
  \includegraphics[width=0.5\textwidth]{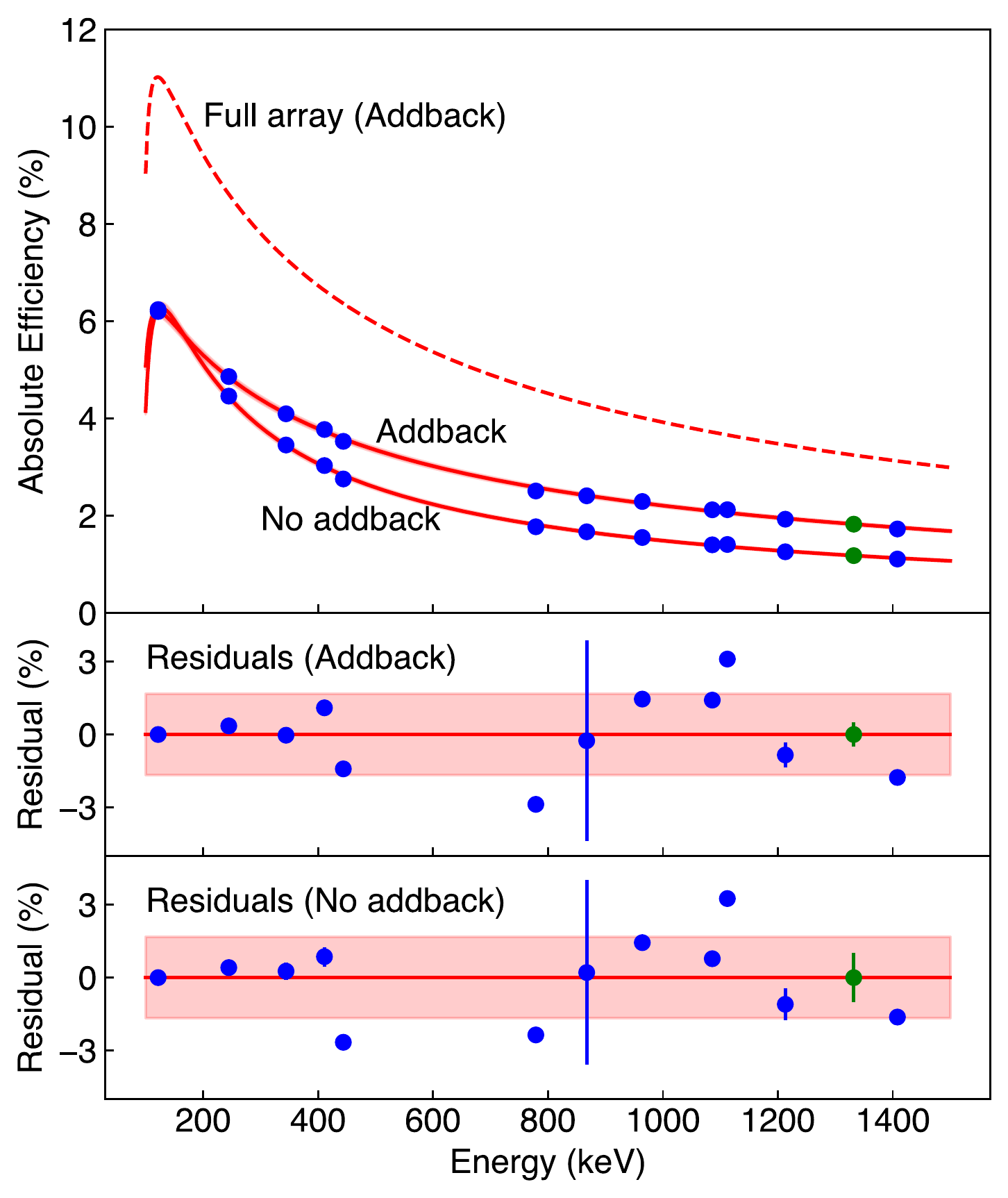}
  \caption{Absolute efficiency of the CLARION2 array. The red
  solid lines are empirical efficiency curves from $^{152}$Eu and
  $^{60}$Co efficiency calibrations (shown as blue and green points,
  respectively), with and without Clover addback.
  These are for the current array, with 9 out of 16
  possible Clovers installed. The dashed red line is the projected
  full efficiency of the array with 16 Clovers. Lower panels show the
  residuals for addback and no addback fits, with shaded areas
  corresponding to the $\approx 2\%$ total uncertainty.}
  \label{fig:clar_eff}
\end{figure}

The efficiency of CLARION2 with the current nine Clovers has been measured using $^{60}$Co and $^{152}$Eu sources. This has been performed in both ``singles'' mode and ``addback'' mode, running triggerless. Addback includes summing two or more crystals in the same Clover that fire within 200~ns of each other. Two independent methods were used with the $^{60}$Co source to establish the absolute efficiency at $1332$~keV: first using the sum-peak at $1173 + 1332 = 2505$~keV relative to the 1173-keV photopeak yielding 1.833(5)$\%$, and second using the 1173-1332 coincidence peak in two different detectors relative to the 1173-keV photopeak yielding 1.815(2)$\%$; both were corrected for the angular correlation. These two methods are independent of the source activity, see Ref.~\cite{Lauritsen2016} for details. The unweighted average of 1.824(9)$\%$ was adopted as the absolute addback efficiency at $1332$ keV, with the spread between the two methods providing a measure of systematic uncertainty. The count rate during the $^{60}$Co measurement was $\approx 100$~Hz per crystal. The relative efficiency was measured using the $^{152}$Eu source, and a smooth function was fitted. This relative efficiency curve was normalised to the absolute efficiency at $1332$~keV extracted from the $^{60}$Co measurement. The efficiency curves for the $^{152}$Eu and $^{60}$Co data points, with and without addback, and residuals from the efficiency fit are shown in Fig.~\ref{fig:clar_eff}. The standard deviation in the residuals was used as an indication of systematic uncertainty, and the shaded region corresponds to the final adopted $\approx 2\%$ uncertainty associated with the absolute efficiency, where summing gains and losses are $<1\%$. The addback factor at 1332-keV is $\epsilon(\text{Addback})/\epsilon(\text{Singles}) = 1.55(2)$. The peak-to-total ratio ($P/T$) for the nine-Clover system with addback and the TRINITY detector installed was measured from the $^{60}$Co source as 0.39 with Compton suppression, and 0.23 without. Without addback, the peak-to-total ratios are 0.20 with Compton suppression, and 0.12 without.

\section{Data Acquisition}
The entire CLARION2-TRINITY array is instrumented with 100- and 500-MHz (14 bit) PIXIE16 waveform digitizers by XIA~\cite{Pixie16}, which enable pulse-shape discrimination, fast timing, timestamps, triggerless operation, and pileup correction in a small form factor. The acquisition system is capable of sustaining 90-MB/s readout for the single crate, which corresponds to roughly 7~TB/day. The system is nominally operated in triggerless mode. Unlike analog systems, which are dead or blocked during the analog-to-digital conversion process, the PIXIE16 digitizers, which buffer list-mode data, allow effectively zero dead time acquisition. However, there is an energy filter dead time associated with the integration and differentiation length of the slow trapezoidal filter, which is typically $<7$~$\mu$s for any single channel. The combination of absolute timestamped events with both fast and slow trapezoidal filters within the PIXIE16 system permits an accurate correction of any pileup or correlated noise that falls within this dead time. This permits accurate absolute counting, see Ref.~\cite{Pixie16} for a more detailed description of the system.

\section{Demonstration of performance}
\begin{figure}
  \centering
  \includegraphics[width=0.5\textwidth]{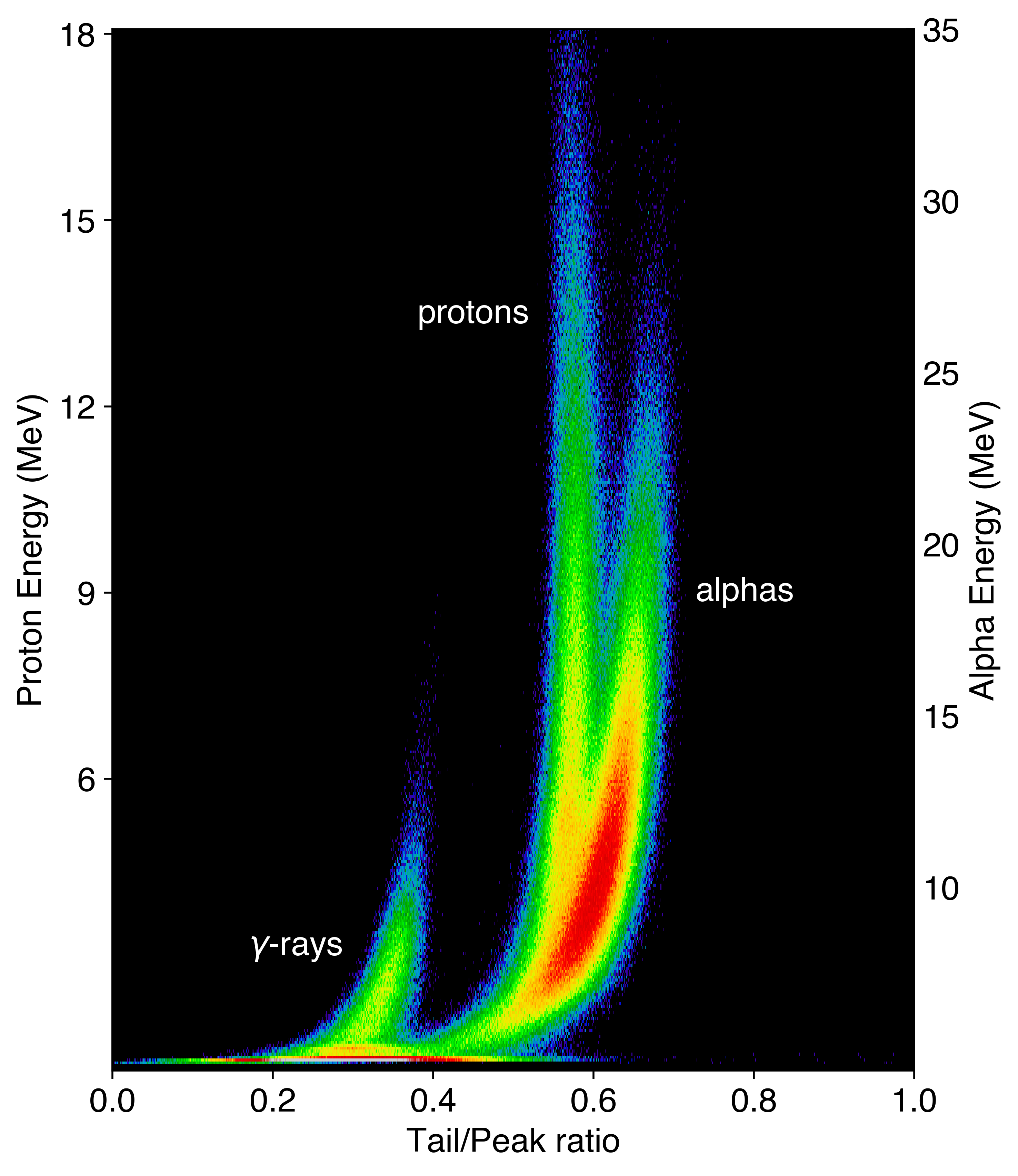}
  \caption{GAGG:Ce PID spectrum from \mbox{$^{16}$O + $^{18}$O}
    fusion-evaporation.}
  \label{fig:fusevap}
\end{figure}

\begin{figure}
  \centering
  \includegraphics[width=0.5\textwidth]{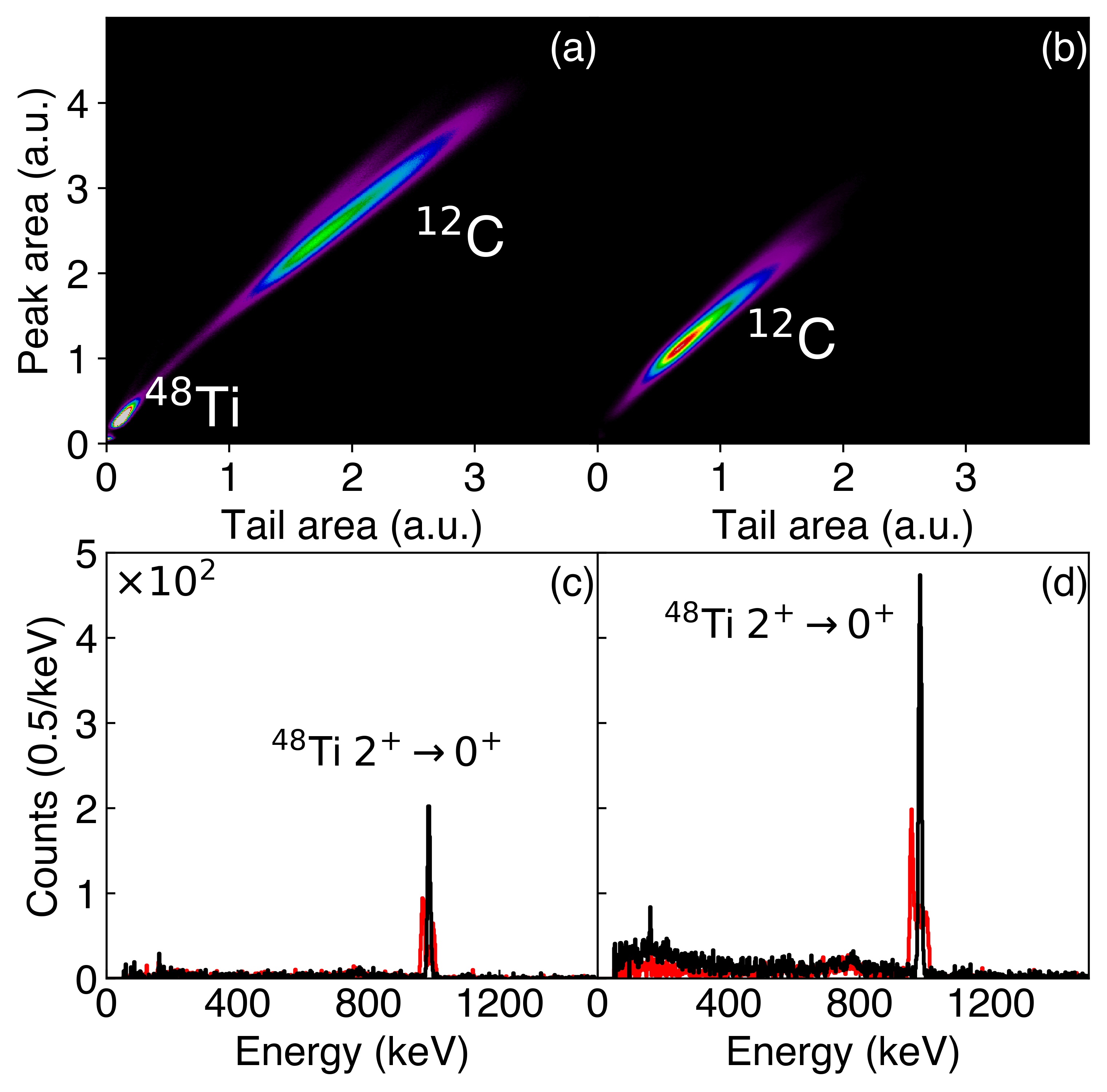}
  \caption{GAGG:Ce PID and Doppler-corrected HPGe spectra from Coulomb excitation of $^{48}$Ti on $^{12}$C. PID spectra are shown in (a) and (b) for Rings 2 and 4, respectively. Doppler-corrected $\gamma$-ray spectra are shown in (c) and (d), for Rings 2 and 4, respectively in black, while the spectra before Doppler correction are shown in red. The FWHM of the 983-keV $2^{+} \rightarrow 0^{+}$ peak is 57~keV before Doppler correction, and 7.8~keV after. The resolution of the Doppler-corrected spectrum is heavily dependent on the mass of the target and beam, as well as the target thickness.}
  \label{fig:coulex}
\end{figure}

Particle spectra from two commissioning experiments conducted at the FSU 9-MV FN Tandem are shown briefly here. Figure~\ref{fig:fusevap} shows the PID spectra for the $^{16}$O + $^{18}$O fusion-evaporation reaction. While protons and alphas are not completely separable for lower energy, at high particle energies there is sufficient separation to provide channel selectivity. Figure~\ref{fig:coulex} shows the PID spectra from the Coulomb excitation of 74.7-MeV $^{48}$Ti on a 0.6-mg/cm$^{2}$ $^{12}$C target. Ring 2 is shown in Fig.~\ref{fig:coulex} (a) and (c) --- both $^{12}$C and $^{48}$Ti recoils are seen in the PID spectrum, while Ring 4 --- Fig.~\ref{fig:coulex} (b) and (d) --- show only the $^{12}$C target recoils. The detected particles were used to Doppler-correct the $\gamma$-rays emitted from the $^{48}$Ti projectiles; the result of this correction is shown in panels (c) and (d), where the $\gamma$-ray peak is sharp and pseudo-Gaussian in shape.

\section{Conclusion}
The design and performance of a new Compton-suppressed HPGe and charged-particle array, CLARION2-TRINITY, have been presented. This new detector system was designed for absolute cross-section measurements with inverse-kinematics reactions, e.g., single-step Coulomb excitation and sub-barrier transfer. However, it also has the flexibility for multi-step Coulomb-excitation and fusion-evaporation measurements, amongst other possibilities. In the course of this development, fundamental characteristics of GAGG:Ce scintillator were determined, including light- and heavy-ion particle identification (PID) capability, pulse-height defects, radiation hardness, and emission spectra. Commissioning experiments have demonstrated the real-world performance of the array, including clean selectivity to the Rutherford scattering in Coulomb excitation of $^{48}$Ti on $^{12}$C. The system in envisioned to primarily operate at the John D. Fox Laboratory of Florida State University, the nuCARIBU-ATLAS facility of Argonne National Laboratory, and the ReA3 facility of FRIB-MSU.

\section*{Acknowledgements}
The authors thank D.~C.~Radford, A.~Galindo-Uribarri, R.~Grzywacz, and W.~Reviol for enlightening conversations on detectors and spectroscopy techniques which stimulated and inspired some of the CLARION2-TRINITY design. This manuscript has been authored by UT-Battelle, LLC under Contract No. DE-AC05-00OR22725 with the U.S. Department of Energy. This work was supported in part by the U.S. National Science Foundation through grant NSF20-12522, by Australian Research Council Grant No. DP170101673, and by the International Technology Center Pacific (ITC-PAC) under Contract Nos. FA520917Q0070 and FA520919PA138. Support for the ANU Heavy Ion Accelerator Facility operations through the Australian National Collaborative Research Infrastructure Strategy (NCRIS) is also acknowledged. T.~J.~G. and M.~S.~M.~G. acknowledge the support of the Australian Government Research Training Program. The publisher acknowledges the US government license to provide public access under the DOE Public Access Plan (http://energy.gov/downloads/doe-public-access-plan).

\bibliographystyle{elsarticle-num}
\bibliography{TrinityNIM}

\end{document}